\documentclass[aps,prl,twocolumn,superscriptaddress,longbibliography,nobalancelastpage]{revtex4-1}

\usepackage{graphicx}
\usepackage{amsmath}
\usepackage{color}
\usepackage[normalem]{ulem}
\usepackage[utf8]{inputenc}
\usepackage{hyperref}

\newcommand{\re}[1]{\text{Re}\left[#1\right]}
\newcommand{\im}[1]{\text{Im}\left[#1\right]}
\newcommand{\eV}{\text{eV}}
\newcommand{\ket}[1]{\left| #1\right\rangle}
\newcommand{\bra}[1]{\left\langle #1\right|}
\newcommand{\Nph}{N_{\text{modes}}}
\newcommand{\Nm}{N_{\text{mol.}}}

\begin{document}
\title{Multimode Organic Polariton Lasing}

\author{Kristin B. Arnardottir}
\email{kba3@st-andrews.ac.uk}
\affiliation{SUPA, School of Physics and Astronomy, University of St Andrews, St Andrews, KY16 9SS, United Kingdom}

\author{Antti J. Moilanen}
\affiliation{Department of Applied Physics, Aalto University School of Science,
P.O. Box 15100, Aalto, FI-00076, Finland}

\author{Artem Strashko}
\affiliation{Center for Computational Quantum Physics, Flatiron Institute, 162 5th Avenue, New York, NY 10010, USA}

\author{Päivi Törmä}
\affiliation{Department of Applied Physics, Aalto University School of Science,
P.O. Box 15100, Aalto, FI-00076, Finland}

\author{Jonathan Keeling}
\affiliation{SUPA, School of Physics and Astronomy, University of St Andrews, St Andrews, KY16 9SS, United Kingdom}

\date{\today}

\begin{abstract}
We present a beyond-mean-field approach to predict the nature of organic polariton lasing, accounting for all relevant photon modes in a planar microcavity.  Starting from a microscopic picture, we show how lasing can switch between polaritonic states resonant with the maximal gain, and those at the bottom of the polariton dispersion.  We show how the population of non-lasing modes can be found, and by using two-time correlations, we show how the photoluminescence spectrum (of both lasing and non-lasing modes) evolves with pumping and coupling strength, confirming recent experimental work on the origin of blueshift for polariton lasing.
\end{abstract}

\maketitle

% Introduction
By placing optically active organic material in a planar microcavity, one can create strong light-matter coupling, and thus new quasi-particles, exciton-polaritons~\cite{Agranovich2009a}.  As seen in many materials~\cite{Carusotto2013a,Sanvitto2016}, when pumped sufficiently, such polaritons transition to a ``condensed'' or lasing state, with macroscopic mode occupation and long range coherence.  A wide variety of organic materials have shown polariton lasing~\cite{kena2010room,Plumhof14,Daskalakis2014,dietrich16,cookson2017,Lerario2017,Ramezani17,scafirimuto2018,Rajendran2019,Wei2019,Vakevainen2019,yagafarov2020blueshift} (for a review, see~\cite{keeling2020review}).  However, there are no general design rules for the optimal material properties for polariton lasing.  Some key ideas have been identified through effective rate-equation modeling~\cite{Michetti2009,Fontanesi2009,Mazza2009,mazza2013microscopic,coles2011vibrationally}, showing how resonance with vibrational modes can play a key role in scattering from an excitonic reservoir to the polariton modes.  However, such effective models leave open many questions, about weak-to-strong coupling crossover, the evolution of coherence and lineshapes, or the competition between lasing modes.

To answer the above questions requires a microscopic model and, as we describe below,  approaches beyond mean-field theory (MFT).
Such mean-field, or single-mode, approaches have been a popular, and powerful, tool~\cite{Cwik2014,Galego15,Herrera2016a,delPino2018:tensorA,delPino2018:tensorB}. However, because they only describe the macroscopically occupied modes, they cannot answer questions about thermalization of the non-lasing modes, nor can they fully describe the photoluminescence profile. For textbook weak-coupling lasers this is well known: MFT considers only coherent photons and stimulated emission, while the semiclassical theory of lasing~\cite{haken70,Scully1997} includes spontaneous emission, which smooths the transition. Here we extend this to describe strong coupling with organic molecules, using a microscopic Hamiltonian model~\cite{Cwik2014,Strashko2018}.  By considering the role of spontaneous processes, we capture fluctuation corrections to MFT, analogous to work on equilibrium excitonic condensates~\cite{Nozieres1985,randeria,Keeling2004,Keeling2005}.  Here we show how to extend this to incoherently pumped and decaying systems.

In this Letter, we develop a second-order cumulant approach to describe the behavior of an organic polariton condensate. Using this we study the evolution of the system with pump strength and cavity detuning. We find a variety of different types of behavior, with lasing either near the bottom of the polariton dispersion, or near resonance with the peak of the gain spectrum. By calculating two-time correlation functions, we present also a microscopic picture of how the polariton dispersion evolves with increasing pumping, giving direct predictions on the polariton blueshift.

\begin{figure}[tpb]
    \centering
    \includegraphics[width=\linewidth]{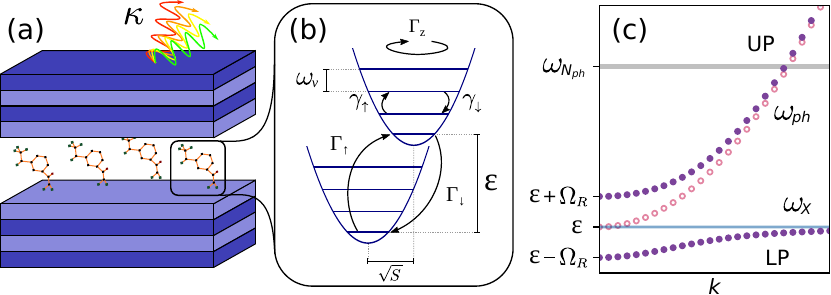}
    \caption{(a) Organic molecules are placed inside a microcavity which supports multiple photon modes. (b) Molecular level structure and processes:
    Incoherent electronic pumping has rate $\Gamma_\uparrow$, electronic decay $\Gamma_\downarrow$, and dephasing $\Gamma_z$. The thermal (de)excitation rates of the vibrational modes are $\gamma_{\uparrow (\downarrow)}$. 
    (c) Discretized photon dispersion. Integer $k$ labels modes. Light-matter coupling hybridizes the photon ($\omega_k$) and exciton ($\varepsilon$) into polariton modes.     Figure plotted for $\omega_0 = \varepsilon$.  We truncate the photon modes at $k=\Nph$ where the hybridization becomes weak.}
    \label{fig:sketch}
\end{figure}

% Model
Our model of organic molecules in multimode planar microcavities is illustrated in Fig.~\ref{fig:sketch}. Following~\cite{Cwik2014,Strashko2018} we model the $\Nm$  molecules as vibrationally dressed emitters, placed randomly in a planar cavity, using an extended \emph{multimode} Tavis--Cummings--Holstein Hamiltonian (i.e. in the rotating wave approximation):
\begin{align}
    H =& \sum_n \left[\frac{\varepsilon}{2}\sigma^z_n + \omega_v\left(b_n^\dagger b_n + \sqrt{S}\left(b_n^\dagger + b_n\right)\sigma_n^z\right)\right]\nonumber\\
    & + \sum_{\mathbf k} \omega_{\mathbf k} a^\dagger_{\mathbf k} a_{\mathbf k} + \sum_{n,{\mathbf k}} \left(g_{n,{\mathbf k}}a_{\mathbf k} \sigma_n^+ + g_{n,{\mathbf k}}^*a_{\mathbf k}^\dagger \sigma_n^-\right).
    \label{eq:Hamiltonian}
\end{align}
The first line describes the organic molecules, labeled by site $n$. Here $\varepsilon$ is the energy of the electronic transition, the Pauli matrices $\sigma_n^{z,\pm}$ describe the electronic state, while $b_n^\dagger$ creates a vibrational excitation (vibron) of energy $\omega_v$. The vibronic coupling is characterized by the Huang--Rhys parameter $S$. 
For brevity we use below the notation $(n-m)$ to denote a transition between the electronic ground state with $n$ vibrons, and the electronic excited state with $m$ vibrons.
The photon modes are labeled by integer $k$, with $a_k^\dagger$ creating a photon in the mode with energy $\omega_k$, and the light-matter coupling has strength $g_{n,k}$; these are discussed in detail below.

We consider photons confined to a two-dimensional planar cavity.  To reduce the problem size, we focus on radially symmetric solutions, label photon modes only through the modulus of their in-plane momentum, and assign a degeneracy factor $W_k$ to account for the two-dimensional density of states. As such, the integer $k$ denotes a plane wave, with wave-vector modulus ${2\pi k}/{L}$,  where $L$ denotes the lateral size of the system. 
The corresponding degeneracies are found by counting the number of lattice  points after dividing up phase space into annuli between radii $k-1/2$ and $k+1/2$ (for $k>0$), along with a disk of radius $1/2$ for $k=0$.  The coupling constants $g_{n,k}$ then take the form $g_{n,k}=g e^{-2i\pi k r_n/L}$, where $r_n$ is the location of molecule $n$, and we choose $g$ to be real~\footnote{We assume molecules are confined to a two-dimensional plane, so there is no dependence on position along the cavity axis.}. The coupling strength depends on the photon mode volume such that $g\sim{1}/{\sqrt{L^2}},$ but as each photon mode couples to many molecules the overall light-matter coupling strength is better characterized by the Rabi splitting $\Omega_R = \sqrt{\Nm}g,$ which will depend on the molecular density in the cavity plane $\rho_{2D} = \Nm/L^2$ but not the size of the system.

The photon mode energies are
$\omega_k = \omega_0 + {E_\rho}k^2/{\Nm}$,
where we have defined an energy scale $E_\rho =  \pi^2 \rho_\text{2D}{\hbar^2}/{2 m}$ in terms of the molecular density $\rho_{2D}$ and effective photon mass $m$. The integer $k$ will have an upper bound, as only those photon modes close enough to resonance with the molecules are resonant.  Taking this condition to be $(\omega_{k_{\text{max}}}-\varepsilon)\gtrsim \Omega_R$  leads to a value $k_{\text{max}} \sim \sqrt{\Nm \Omega_R/E_\rho}$.  Counting the total number of photon modes, $\Nph$, in a 2D system, $\Nph \propto k_{\text{max}}^2$, so we find 
${\Nph}/{\Nm}\sim{\Omega_R}/{E_\rho}$. 
This ratio is important, as it determines how well MFT works. For a single mode, MFT is controlled by the $1/\Nm$ suppression of fluctuations. As we discuss below, adding more photon modes increases fluctuations, so that the relevant ratio is $\Nph/\Nm$.
For realistic systems $E_\rho$ is typically of the order $10^5-10^6$ eV~\cite{eizner2019inverting}, far greater than the typical value of $\Omega_R\simeq 1$eV. For the numerical results in this paper we have set $E_\rho=5\times 10^{5}$eV.

Including the incoherent processes shown in Fig.~\ref{fig:sketch}, the equation of motion for the system density matrix is~\cite{Strashko2018}:
\begin{multline}
    \label{eq:Lindblad}
    \partial_t \rho =  -i[H,\rho] + \sum_k \kappa \mathcal{L}[a_k]
        + \sum_n\Big(\Gamma_\uparrow\mathcal{L}[\sigma^+_n] 
        + \Gamma_\downarrow\mathcal{L}[\sigma^-_n]\\
         + \Gamma_z \mathcal{L}[\sigma^z_n]
        + \gamma_\uparrow \mathcal{L}[b^\dagger_n - \sqrt{S}\sigma^z_n]
        + \gamma_\downarrow \mathcal{L}[b_n - \sqrt{S}\sigma^z_n]
            \Big),
\end{multline}
% Rewrite bellow
where $\mathcal{L}[X] = X\rho X^\dagger - \frac{1}{2}(X^\dagger X \rho + \rho X^\dagger X)$. We include photon loss at rate $\kappa$ (assumed equal for all  modes), and incoherent pumping,  decay, and dephasing of molecules with rates $\Gamma_\uparrow$, $\Gamma_\downarrow$, and $\Gamma_z$ respectively. The last two terms describe thermalization of vibrons, with rates $\gamma_\downarrow = \gamma_v (n_b+1)$, $\gamma_\uparrow = \gamma_v n_b $ and
$n_b = \left[\text{exp}(\omega_v/k_b T_v) - 1\right]^{-1}$ is the Bose--Einstein occupation at temperature $T_v$.  Physically this corresponds to the assumption that vibrational modes of the molecules are rapidly thermalized by a bath of delocalized phonon modes. 

To capture both strong vibrational and light-matter coupling, we will combine  two electronic and $N$ vibrational levels into $2N$-level molecular operators for each molecule. A basis for such operators are the generalized Gell-Mann matrices~\cite{stone2009mathematics}, $\lambda_i^{(n)}$. Using these, the system Hamiltonian takes the form: 
$H = \sum_k \omega_k a_k^\dagger a_k + \sum_n \big[A_i + \sum_k(B_i a^\dagger_k e^{-2i\pi kr_n/L} + h.c.)\big]\lambda_i^{(n)}$.
Hereon, sums over repeated Gell-Mann matrix indices $i$ are implicit. The form of the vectors $A_i$ and $B_i$ is determined by Eq.~\eqref{eq:Hamiltonian}. Equation \eqref{eq:Lindblad} can similarly be rewritten: $\partial_t \rho = -i[H,\rho] +  \sum_k \kappa \mathcal{L}[a_k] +\sum_{\mu,n} \mathcal{L}\left[\gamma_i^{\mu} \lambda^{(n)}_i\right]$. Here $\mu$ runs over the different molecular dissipative processes (pump, decay, dephasing, and vibrational excitation and decay).

%Look at section bellow. Might be moved partially to Intro
As noted above, mean-field (MF) approaches are useful to understand the linear stability of the normal state, but cannot yield information about the non-lasing modes, which ultimately can modify the critical properties of the lasing transition.  In thermal equilibrium, this was studied by considering Gaussian fluctuation corrections to MFT~\cite{Nozieres1985,randeria,Keeling2004,Keeling2005}.  There one finds that when fluctuation corrections are large, the critical temperature matches the degeneracy temperature of the 2D Bose gas, due to thermal depopulation of the condensate mode.  The relative significance of fluctuations in those works depends on a dimensionless ratio, $m^\ast = m \Omega_R/\hbar^2 \rho_{2D}$ --- meaning fluctuations are more significant when there is a large photon density of states.  This quantity $m^\ast$ is the same ratio as  ${\Nph}/{\Nm}$, identified as controlling the role of beyond-mean-field effects. We will see that even in the general non-equilibrium context, this same parameter controls the effects of fluctuations.  Given the small value we have for $\Nph/\Nm$, we may expect the effects of fluctuations will be small, but non-vanishing.  A further discussion of the effects of changing this parameter is given in the supplemental material~\cite{SM}.

To go beyond MFT, we write second order cumulant equations; this means writing equations of motion for second-order correlations of the operators $a^{}_k,a^\dagger_k$ and $\lambda_i^{(n)}$, and splitting all higher order expectations into products of first and second order moments~\cite{gardiner2009stochastic}. 
Such an approach directly relates to the semiclassical theory of lasing~\cite{haken70}, and has been used to study the differences between lasing and condensation in the Dicke model~\cite{Kirton2017,Kirton2018}, and to study Rabi oscillations~\cite{SanchezBarquilla2020}.  Here we extend this approach to describe competition between different photon modes.
As the Hamiltonian has $U(1)$ symmetry under a common phase change of $a_k$ and $\sigma^-_n,$ the cumulant equations simplify if we split the $\lambda_i$ into three groups: those which conserve, increase or decrease the electronic excitations, denoted  $z, +,$ and $-$ respectively.
We also assume spatial homogeneity, which leads to conservation of momentum, eliminating cross-$k$ terms.  For single-mode lasing, this also leads to homogeneous populations, so that $\ell_i=\langle \lambda_{i_z}^{(n)}\rangle$ is independent of index $n$. For coherences, we similarly can write the non-zero Fourier components as: $c_i^k=\sum_n e^{i2\pi r_n k/L }\langle a^{}_k \lambda_{i_+}^{(n)}\rangle/\Nm$ and $d_{ij}^k=\sum_{n,m\neq n} e^{i2\pi (r_n-r_m) k/L }\langle \lambda_{i_+}^{(n)}\lambda_{i_-}^{(m)}\rangle/\Nm^2.$ 
Note that limiting to these expressions implies there is no coherence between different $k$ modes --- this is equivalent to the assumption of homogeneous populations.
The equations of motion for these quantities, along with the photon occupations $n^k$ take the form:
\begin{align}
\label{eq:eomn}
\partial_t n^k =& -\kappa n^k - 2\Nm \im{B_i c_i^k}, \\
\label{eq:eoml}
\partial_t \ell_i =& \xi_{ij} \ell_j +\phi_i - 8 \re{\beta_{ij} \sum_k c_j^k}, \\
\label{eq:eomc}
\partial_t c_i^k =& \left[X_{ij} -\left(i\omega_k+\frac{\kappa}{2}\right)\delta_{ij}\right]c_j^k + 2\beta^*_{ij}\ell_j n_k \nonumber\\
    &-iB_j d_{ij}^k -\frac{i}{\Nm}\left(\zeta_{ij}\ell_j + \frac{B_i}{2N} \right), \\
\label{eq:eomd}
\partial_t d_{ij}^k =& X^*_{ip} d^k_{pj} + X_{jp}d^k_{ip} +2\ell_p\left(\beta_{ip}\Tilde{c_j}^k + \beta^*_{jp}\Tilde{c_i}^k\right),
%    & +2\ell_p\left[\beta_{ip}\left(c_j^k-\frac{1}{\Nm}\sum_{k'} c_j^{k'}\right) + \beta^*_{jp}\Tilde{c_i}^k\right]
\end{align}
where $\Tilde{c_i}^k = c_i^k -\sum_q c_i^q/\Nm$ and the coefficients $\xi_{ij}, X_{ij}, \phi_i, \beta_{ij}$, and $\zeta_{ij}$ are different combinations of the relevant structure constants, Hamiltonian constants and decay rates, specified in the supplemental material~\cite{SM}.  By going beyond MF approaches, these equations now enable us to correctly capture the multimode behavior, and see how fluctuations modify the lasing threshold.

\begin{figure}
    \centering
    \includegraphics[width=\linewidth]{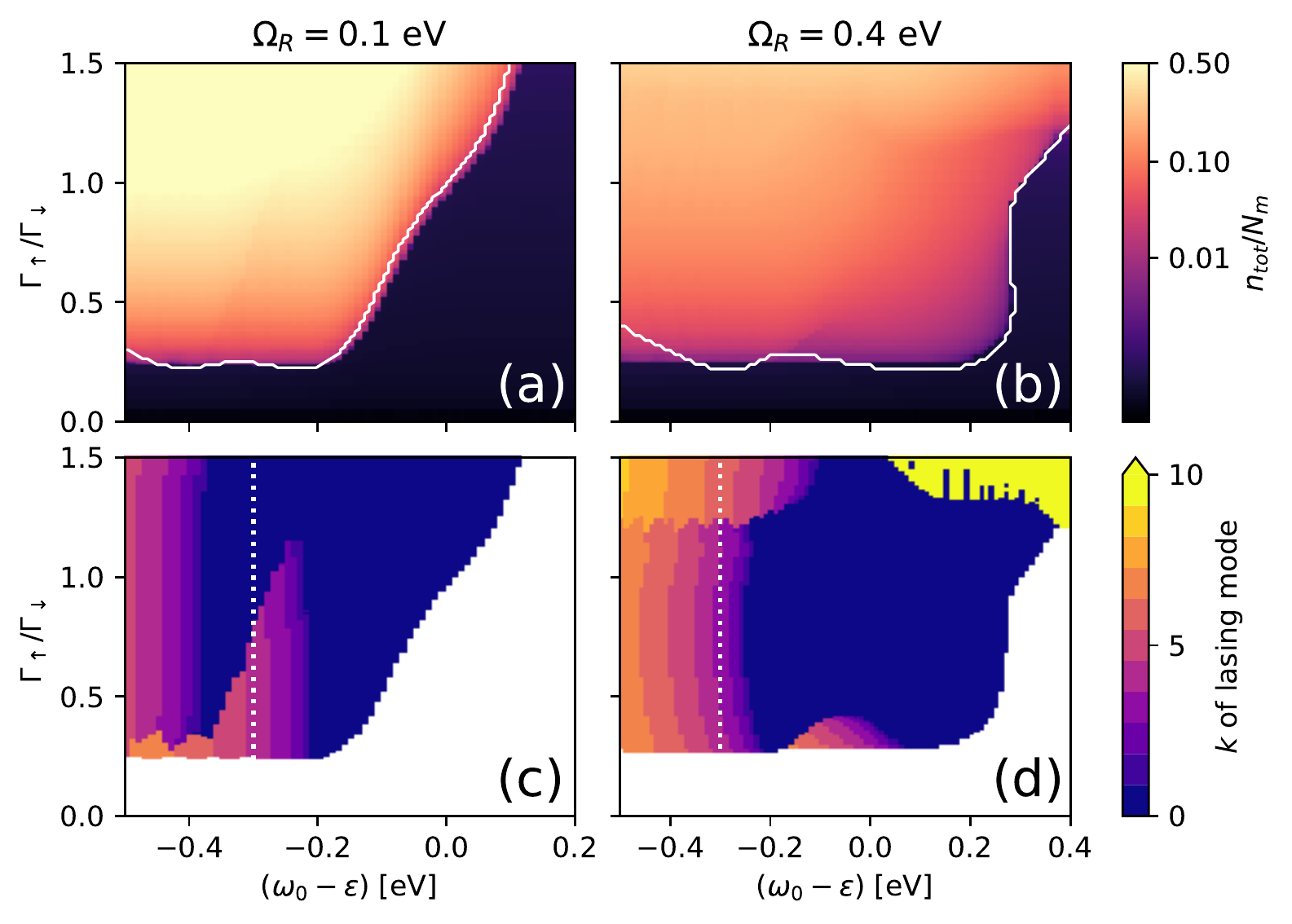}
    \caption{Phase diagrams for the system at two values of the Rabi splitting $\Omega_R$.
    The top row shows total photon occupation vs external pump rate (vertical) and the detuning of the photon dispersion from the zero vibron transition (horizontal). The white line is the single mode, MF linear stability result. 
    The bottom row shows the $k$ index of the highest occupied photon mode. The white dashed line indicate the parameters used in Fig. \ref{fig:modeswitching}(a-b).
    Parameters used here: $S=0.1, \omega_v=0.2 \eV, \Gamma_\downarrow = \kappa = 10^{-4} \eV, \Gamma_z= 0.03 \eV, \gamma_v = 0.02 \eV, k_bT_v= 0.025\eV$}
    \label{fig:phasediagrams}
\end{figure}

Figure~\ref{fig:phasediagrams}(a-b) shows phase diagrams of total photon occupation, $n_{tot} = \sum_k W_k n^k$, vs photon detuning $\omega_0$ and pump $\Gamma_\uparrow$. 
Where MF predicts lasing (white line, from Ref.~\cite{Strashko2018}) there is macroscopic occupation, while outside that region there is only occupation of order one.   The agreement between the MF and cumulant equations holds only because $\Nm\gg\Nph$; in~\cite{SM} we show that for smaller $\Nm$, fluctuations push the transition to higher pump power.
The only exception is in the bottom left corner,  for values of $\omega_0$ where MF predicts no lasing, but there exist a larger $\omega_0$ which would show lasing at the same pump strength. In this case, the multimode model permits lasing of  $k>0$ modes.
We find lasing at non-zero  $k$ occurs in other cases as well. Figures~\ref{fig:phasediagrams}(c-d) show which photon mode has the largest occupation $n_k$ in the lasing region.
In the case of weaker coupling,($\Omega_R=0.1$eV, left column) we find that when $\omega_0$ is close to $(n-0)$ transitions (i.e. for $\omega_0=\varepsilon-n\omega_v$) for $n=0,1,2$,  lasing is predominantly in the $k=0$ mode. For $\omega_0$ between the $(1-0)$ and $(2-0)$ transitions there is competition between $k=0$ and the mode, denoted $k^\ast$ below, closest to resonance with the $(1-0)$ transition. 

\begin{figure}
    \centering
    \includegraphics[width=\linewidth]{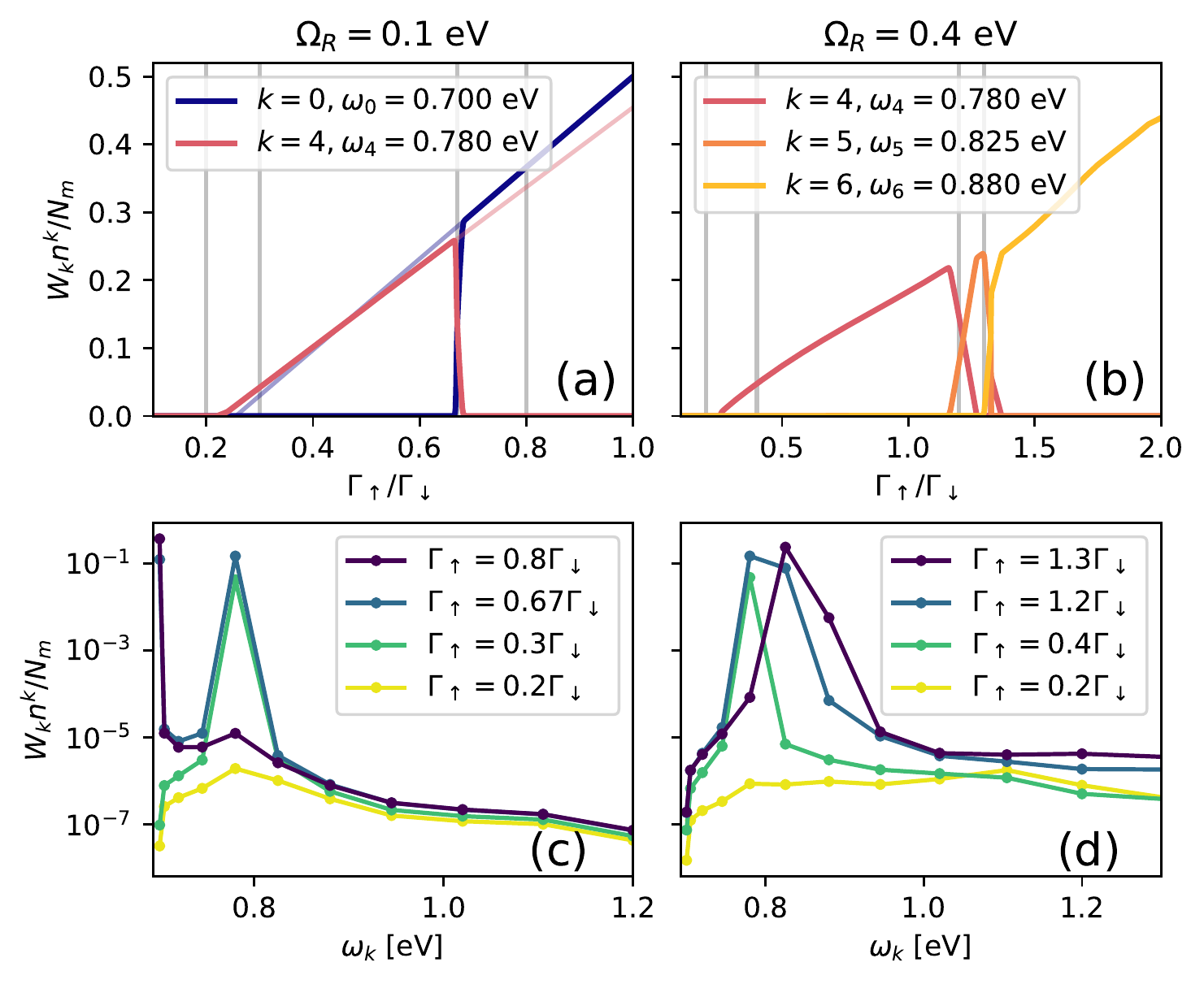}
    \caption{Lasing mode switching for two different Rabi splittings $\Omega_R$. The top row shows photon occupation as a function of external pump when $\omega_0$ falls in between the $(1-0)$ and $(2-0)$ transitions ($\omega_0-\varepsilon = -1.5\omega_v$). (a) For Rabi splitting of $\Omega_R=0.1$eV, the non-zero vibron transitions (vibronic sidebands) are outside of the range of strong coupling. There is clear switching between two lasing modes, one close to $(1-0)$ and the other at $k=0$. Each mode lases it behaves linearly (a linear fit is shown by the thinner, fainter lines), similarly to textbook weak-coupling single-mode models. Ref.~\cite{SM} shows the threshold behavior on logarithmic scale.  
    (b) For Rabi splitting of $\Omega_R=0.4$eV the lasing behavior is strongly influenced by the light matter coupling, and the curves are no longer linear.
    The bottom row shows the occupation of all photon modes vs bare photon frequency $\omega_k$, for the same values of $\omega_0$ and $\Omega_R$ as in (a,b), and pump strengths as marked by the gray vertical lines in panels (a,b).
    Other parameters are the same as in Fig.~\ref{fig:phasediagrams}. }
    \label{fig:modeswitching}
\end{figure}

To better see this competition, Fig.~\ref{fig:modeswitching} shows $n_k$ as a function of the pump for $\omega_0 = \varepsilon -1.5\omega_v = 0.7$ eV. For weaker light-matter coupling, the growth of photon occupation is linear with pumping. The mode switching can then be explained by the competition of two potential lasing modes: the lowest mode $k=0$ and  a non-zero $k$ mode, resonant with the $(1-0)$ transition. Each mode has its own threshold and slope efficiency (gradient of photon occupation vs  pump). When two modes are above threshold, the mode with larger gain will suppress the other. There is a very narrow region of pump values where the gain is similar, which leads to coexistence~\footnote{The assumption of single-mode operation used in Eq.~(\ref{eq:eomn}--\ref{eq:eomd}) will break down in this coexistence region, however this region is very narrow, so we do not expect significant deviations to arise.} --- macroscopic occupation of both modes --- this can be seen from e.g. $\Gamma_\uparrow=0.67\Gamma_\downarrow$ line in Fig.\ref{fig:modeswitching}(c). 
Another notable feature is that even at the lowest detunings, lasing never switches to a mode resonant with the $(3-0)$ transition. This can explained by the coupling to this transition being too weak, as the effective coupling  to the $(n-0)$ transitions falls off as $\langle n|e^{\sqrt{S}(b^\dagger-b)}|0\rangle= S^{n/2}/\sqrt{n!}$.

At stronger light-matter coupling, as shown in the right columns of Figs.~\ref{fig:phasediagrams}, \ref{fig:modeswitching}, the mode switching can no longer be described as a patchwork of single mode results. Figure~\ref{fig:phasediagrams}(d) shows two interesting changes. First, there is a region of high $k^\ast$ lasing for very high pump strength and positive detuning, and secondly at large negative detuning there is switching between adjacent $k^\ast$ modes. The high $k^\ast$ lasing can be explained by the complete inversion of the two-level system, which implies net gain exists at both high and low frequencies (away from the vibronic structure);  photon modes at high $k^\ast$ overlap with such gain.
A large positive $\omega_0$ is required to reach total inversion of the two-level systems, as otherwise, polariton lasing clamps $\langle \sigma^z\rangle<0$. For further details see Ref.~\cite{SM}.
To understand the $k^\ast$ switching we consider in detail the behavior seen in Fig.~\ref{fig:modeswitching}(b).   Because for $\Omega_R=0.4$eV, the system remains in the strong coupling regime even when lasing (see below), we may note that even at a fixed $k^\ast$, the energies of the polaritons are known to shift to higher energy with increasing density (also discussed below).  As such, the shift to higher $k^\ast$ with increasing pump is at first surprising: lasing moves to higher energy modes, as each mode itself moves to higher energy.  The explanation of this requires the observation that with increasing pumping, the gain spectrum also shifts to higher energies.
This happens because the sequence of $(n-0)$ vibrational sidebands, with decreasing $n$ and thus increasing energy, become inverted in turn as pumping increases.

\begin{figure}
    \centering
    \includegraphics[width=\linewidth]{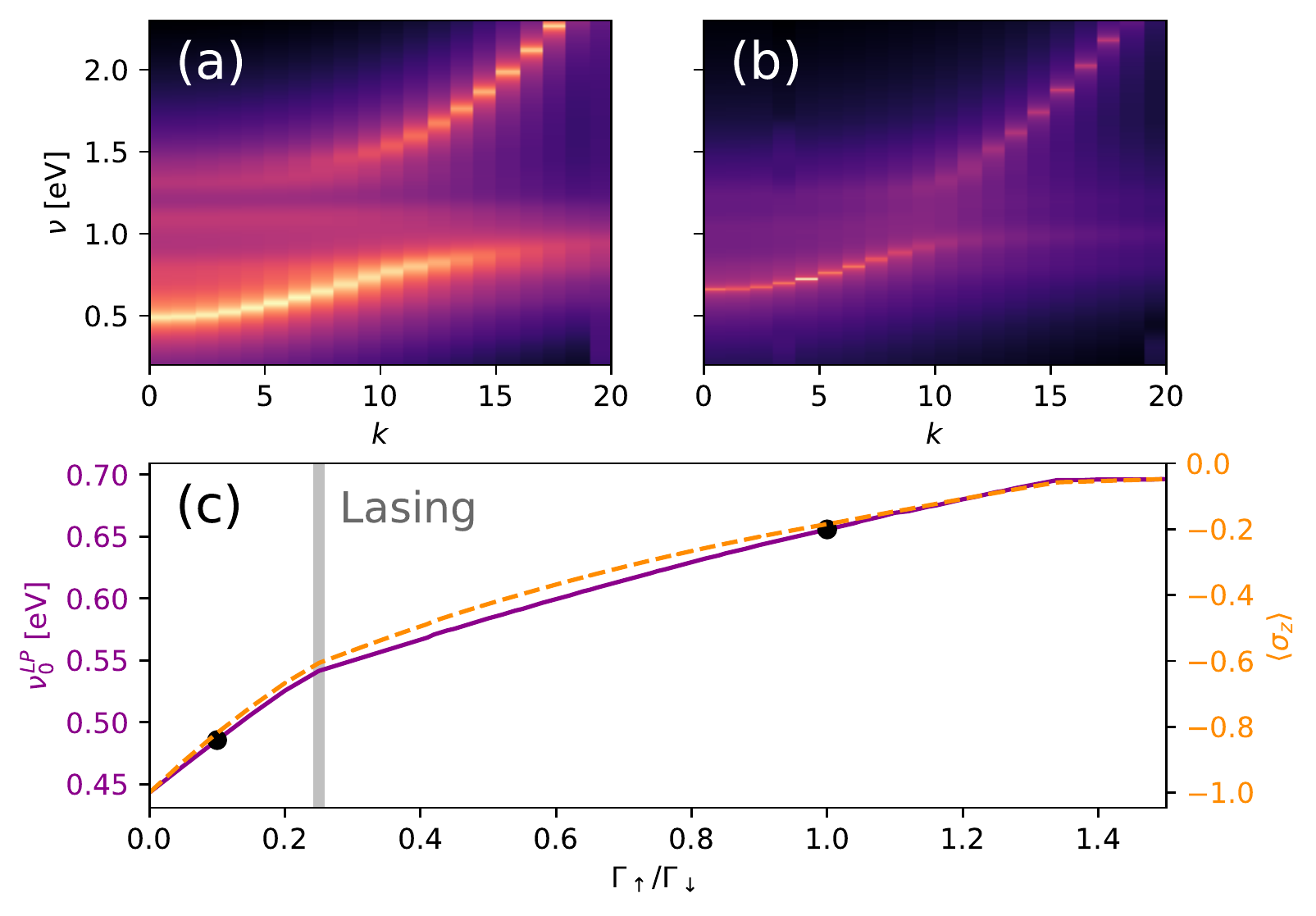}
    \caption{(a-b) Photoluminescence spectra of the system for (a) $\Gamma_\uparrow=0.1\Gamma_\downarrow$ and (b) $\Gamma_\uparrow=\Gamma_\downarrow
    $. (c) The energy of the lower polariton branch $\nu^{LP}_0$ (purple, solid line, left axis) and $\langle \sigma^z\rangle$ (dashed, orange line, right axis) vs pump strength. The black dots correspond to the pump values in panels (a-b) above, and the gray line marks the lasing threshold.}
    \label{fig:blueshift}
\end{figure}

As already noted, due to strong light-matter coupling, the mode energies are those of polaritons, not bare photons and excitons.  Moreover, because of the saturability of two-level systems, these polariton energies are density dependent, showing a blueshift of the lower polariton with increasing pumping.  To find the energy of the occupied modes,  we can calculate the photoluminescence (PL) spectrum:
\begin{equation}
    \label{eq:PL}
    S_k(\nu) = \int_{-\infty}^\infty dt \langle a_k^\dagger (t) a^{}_k(0)\rangle e^{i\nu t}.
\end{equation}
This can be found by the quantum regression theorem~\cite{Scully1997}: Using the steady state density matrix $\rho_{ss}$, we construct $\tilde{\rho}(0) = a_k\rho_{ss}$, time evolve $\tilde{\rho}$, and then evaluate $\text{Tr}\big[ a_k^\dagger \tilde{\rho}(t) \big]$. Within our cumulant approximation, this gives a  set of coupled differential equations for two-time correlators, with a form similar to Eqs.~\eqref{eq:eomn}--\eqref{eq:eomd}, see~\cite{SM}.  

Example PL spectra are shown in Fig.~\ref{fig:blueshift}(a-b), for parameters matching the strong light-matter coupling shown in Fig.~\ref{fig:modeswitching}(b). Panel (a) corresponds to a small pumping while (b) shows larger pumping in the lasing regime. The peaks in the spectra correspond to the system's mode energies, their widths to the lifetimes, and intensities to the occupations. The polariton anti-crossing is clearly visible in both spectra, indicating strong coupling persists, but the splitting is reduced at higher pumping, corresponding to a blueshift of the lower polariton branch. In contrast, for $\Omega_R=0.1$eV, the polariton splitting collapses before lasing occurs.
We may note that even when strong coupling is seen, a broad feature at the bare exciton energy is visible.  This corresponds to uncoupled excitonic ``dark states'', which are known to become optically active due to the vibronic coupling~\cite{cwik2016excitonic,Herrera2016a,zeb2017exact}.
We can study the blueshift by extracting the lower polariton frequency at $k=0$ for different pumping strengths, as seen in Fig.~\ref{fig:blueshift}(c) in solid (purple) --- see \cite{SM} for details. By comparing the lower polariton energy to the inversion $\langle \sigma_z\rangle$ (dashed, orange line) it is clear that the blueshift seen here corresponds entirely to the saturation of the molecular optical transition.

%Conclusion
In this Letter we have shown how the nature of organic polariton lasing  changes with changing pump, detuning, and light-matter coupling.  To understand this fully requires consideration of the multiple photon modes in a planar microcavity, which in turn demands a treatment beyond MFT, which we have introduced here.
We find switching between different lasing modes, which can be understood via the slope efficiencies of different modes, and the evolution of gain profile.
Our approach allows direct calculation of the PL spectrum of the driven system, giving information on the lasing frequency and the evolution of the polariton dispersion,  distinguishing  photon and polariton lasing. Using our microscopic model, we could show that the blueshift closely matches the occupation of the exciton ground state, corroborating the phenomenological saturation model of polariton interaction~\cite{yagafarov2020blueshift}.  The methods described in this Letter can straightforwardly be extended to more complex molecules  (e.g. other electronic states, further vibrational modes, or other dissipative processes), or to analysis of time-dependent pumping which we will discuss in a subsequent publication.  As such, this provides a foundation to predict how molecular properties determine the optimal materials for organic polariton lasing.

{\em Acknowledgments---}
We acknowledge Peter Kirton for helpful comments on an earlier version of this manuscript.
KBA and JK acknowledge financial support from EPSRC program ``Hybrid Polaritonics'' (EP/M025330/1). AJM and PT acknowledge support by the Academy of Finland under project numbers 303351, 307419, 327293, 318987 (QuantERA project RouTe) and 318937
(PROFI), and by Centre for Quantum Engineering (CQE) at Aalto University. AJM acknowledges financial support by the Jenny and Antti Wihuri Foundation. The Flatiron Institute is a division of the Simons Foundation. AS acknowledges support 
from the EPSRC CM-CDT (EP/L015110/1).

\onecolumngrid
\clearpage

\renewcommand{\theequation}{S\arabic{equation}}
\renewcommand{\thefigure}{S\arabic{figure}}
\setcounter{equation}{0}
\setcounter{figure}{0}
\setcounter{page}{1}

\section{Supplemental Material for: ``Multimode Organic Polariton Lasing''}
\twocolumngrid

\section{Form of Cumulant Equations}
This section gives more detail on the derivation of the cumulant equations, Eqs.~(3-6) of the main text. 

It is first important to note some useful identities of the general Gell-Mann matrices (GMM)~\cite{stone2009mathematics}.  These obey:
\begin{displaymath}
[\lambda_i, \lambda_j] = 2if_{ijp} \lambda_p,\qquad
\lambda_i \lambda_j = \zeta_{ijp} \lambda_p + \frac{2}{N_\lambda} \delta_{ij}
\end{displaymath}
with $\zeta_{ijp}=if_{ijp} + t_{ijp}$ and $N_{\lambda}$ as the dimension of the $\lambda$ matrices (which here is $2N$, where $N$ is the number of vibrational levels). The tensors $f_{ijk}$ and $t_{ijk}$ are totally  antisymmetric and symmetric tensors, respectively.

\begin{widetext}
Using these identities, the equations of motion corresponding to the master equation in the main text are:
\begin{align}
\partial_t \langle a^\dagger_k a_k\rangle = & - \kappa \langle a^\dagger_k a_k\rangle - 2\sum_n\text{Im}\left( B_i^* e^{2i\pi k r_n/L} \langle a_k\lambda_i^{(n)}\rangle\right)\label{eq:eom-ph},\\
%%%%%
\partial_t \langle\lambda_i^{(n)}\rangle = & \xi_{ij}\langle \lambda_j^{(n)}\rangle + \phi_i +  4f_{ijp}\sum_k\text{Re}\left( B_j^* e^{2i\pi k r_n/L} \langle a_k\lambda_p^{(n)}\rangle\right)\label{eq:eom-ell},\\
%%%%%%
\partial \langle a_k \lambda_i^{(n)}\rangle = &  \left[ \xi_{ij} - \left(i\omega_k + \frac{\kappa}{2}\right)\delta_{ij}\right] \langle a_k \lambda_j^{(n)}\rangle\nonumber\\
  &	 + 2 f_{ijp}\sum_{k'}
	   \left(B_j e^{-ik'r_n} \langle\lambda_p^{(n)} a_{k'}^\dagger a_k\rangle
	 + B_j^*e^{ik'r_n} \langle\lambda_p^{(n)} a_{k'} a_k\rangle \right)\nonumber\\
	 & - i B_j \left(\zeta_{ijp} e^{-2i\pi k r_n/L} \langle\lambda_p^{(n)}\rangle 
	 +\frac{1}{N} e^{-2i\pi k r_n/L} \delta_{ij}
	 + \sum_{m\neq n} e^{-ikr_m} \langle \lambda_i^{(n)} \lambda_j^{(m)}\rangle
	 	\right)\label{eq:eom-cpl},\\
	 	%%%%%%
\partial_t \langle \lambda_i^{(n)} \lambda_j^{(m)}\rangle = & 
	\xi_{ip} \langle \lambda_p^{(n)} \lambda_j^{(m)}\rangle +
	\xi_{jp} \langle \lambda_i^{(n)} \lambda_p^{(m)}\rangle\nonumber\\
	& +
	4\sum_k\left[f_{ilp}\text{Re}\left( B_l^* e^{2i\pi k r_n/L} 
	\langle a_k\lambda_p^{(n)}\lambda_j^{(m)}\rangle\right)
	+ f_{jlp}\text{Re}\left( B_l^* e^{ikr_m} 
	 \langle a_k\lambda_i^{(n)}\lambda_p^{(m)}\rangle\right)
	\right],\label{eq:eom-d}
\end{align}
\end{widetext}
where we have for brevity defined,
\begin{align*}
\xi_{ij} &= 2 f_{ipj} A_p +  
	i \left(
	f_{ips}\zeta_{rsj} + f_{ris}\zeta_{spj}\right)\sum_\mu \gamma_p^\mu \gamma_r^{\mu *}\\
\phi_i &= \frac{2i}{N}f_{ipr} \sum_\mu \gamma_p^\mu \gamma_r^{\mu *}.
\end{align*}
As a reminder, $\kappa$ is the cavity loss rate, 
$\omega_k$ the cavity mode energies.
The quantities 
$A_i, B_i$ relate to the rewriting of the system Hamiltonian in terms of GMM as  $H = \sum_k \omega_k a_k^\dagger a_k + \sum_n \big[A_i + \sum_k(B_i a^\dagger_k e^{-2i\pi kr_n/L} + h.c.)\big]\lambda_i^{(n)}$.  Similarly $\gamma^\mu_r$ correspond to the dissipation terms written as
$\sum_{\mu,n} \mathcal{L}\left[\sum_i \gamma_i^{\mu} \lambda^{(n)}_i\right]$

We take third order cumulants to be zero, i.e. we assume the correlations between three operators is captured by a combinations of two operator correlators~\cite{gardiner2009stochastic}. We can then rewrite third order terms using
$\langle ABC\rangle = \langle A\rangle \langle BC\rangle
+\langle B\rangle \langle AC\rangle
+\langle C\rangle \langle AB\rangle$. It becomes convenient to distinguish between different types of GGMs: $z$ matrices do not affect the electronic state of the molecule while $x$ ($y$) matrices do affect the electronic state and are real (imaginary). Because of the $U(1)$ symmetry of the Hamiltonian we can look at $x$ and $y$ matrices as odd (along with the photon operators $a^{(\dagger)}$) and any term with an odd number of those operators will average to zero.  

\begin{widetext}
The non-vanishing cumulant equations which survive are the following:
\begin{align}
\partial_t \langle a^\dagger_k a_k\rangle = & - \kappa \langle a^\dagger_k a_k\rangle - 2\sum_n\text{Im}\left( B_i^* e^{2i\pi k r_n/L} \langle a_k\lambda_{i_\alpha}^{(n)}\rangle\right),\\
%%%%%
\partial_t \langle \lambda_{i_z}^{(n)}\rangle = & \xi_{ij}\langle \lambda_{j_z}^{(n)}\rangle + \phi_i +  4f_{ijp}\sum_k\text{Re}\left( B_j^* e^{2i\pi k r_n/L} \langle a_k\lambda_{p_\alpha}^{(n)}\rangle\right),\\
%%%%%%
\partial \langle a_k\lambda_{i_\alpha}^{(n)}\rangle = &  \left[ \xi_{ij} - \left(i\omega_k + \frac{\kappa}{2}\right)\delta_{ij}\right] \langle a_k \lambda_{j_\alpha}^{(n)}\rangle\nonumber\\
  &	 + 2 f_{ijp}\langle\lambda_{p_z}^{(n)}\rangle\sum_{k'}
	   \left(B_j e^{-ik'r_n} \langle a_{k'}^\dagger a_k\rangle
	 + B_j^*e^{ik'r_n} \langle a_{k'} a_k\rangle \right)\nonumber\\
	 & - i B_j \left(\zeta_{ijp} e^{-2i\pi k r_n/L} \langle \lambda_{p_z}^{(n)}\rangle 
	 +\frac{1}{N} e^{-2i\pi k r_n/L} \delta_{ij}
	 + \sum_{m\neq n} e^{-ikr_m} \langle \lambda_{i_\alpha}^{(n)} \ \lambda_{j_\alpha}^{(m)}\rangle
	 	\right),\\
	 	%%%%%%
\partial_t \langle \lambda_{i_\alpha}^{(n)} ~\lambda_{j_\alpha}^{(m)}\rangle = & 
	\xi_{ip} \langle \lambda_{p_\alpha}^{(n)}  \lambda_{j_\alpha}^{(m)}\rangle +
	\xi_{jp} \langle \lambda_{i_\alpha}^{(n)}  \lambda_{p_\alpha}^{(m)}\rangle\nonumber\\
	& +
	4\sum_k\left[f_{ilp}\text{Re}\left( B_l^* e^{2i\pi k r_n/L} 
	\langle a_k\lambda_{p_z}^{(n)}\lambda_{j_\alpha}^{(m)}\rangle\right)
	+ f_{jlp}\text{Re}\left( B_l^* e^{ikr_m} 
	 \langle a_k\lambda_{i_\alpha}^{(n)} \lambda_{p_z}^{(m)}\rangle\right)
	\right].
\end{align}
\end{widetext}
where the notation $\lambda_{i_\alpha}$ means that we only take matrices that corresponds to transitions that change the electronic excitations (proportional to $\sigma^x$ or $\sigma^y$) while $\lambda_{i_z}$ matrices do not change the electronic excitation. To fully take advantage of the symmetry presented by the rotating wave approximation (i.e. the conservation of excitation number) we define new matrices that are proportional to $\sigma^{\pm}: \lambda_{i_\pm}= \frac{1}{2}\left(\lambda_{i_x}\pm \lambda_{i_y}\right).$ We can then define the discrete Fourier transforms
\begin{align}
    c_i^k &= \frac{1}{\Nm}\sum_n e^{i2\pi r_n k/L} \langle a_k \lambda_{i_+}^{(n)}\rangle\\
    d_{ij}^k &=\frac{1}{\Nm^2}\sum_{n,m\neq n} e^{i2\pi (r_n-r_m) k/L} \langle \lambda_{i_+}^{(n)} \lambda_{i_-}^{(m)}\rangle
\end{align}
and arrive at the equations in the main text (Eqs. (\ref{eq:eomn}-\ref{eq:eomd})) with the coefficients
$X_{ij} = \xi_{i_xj_x}+i\xi_{i_yj_x}$ and $\beta_{ij} = B_p(f_{i_x pj}-i f_{i_y pj})/2$.

\section{Scaling of fluctuations with \texorpdfstring{$\Nm$}{Nm}}

As mentioned in the main text, the ratio $\Nph/\Nm$ controls how significant corrections to mean field theory are.  This section presents further results illustrating this dependence.

\begin{figure}[htpb]
    \centering
    \includegraphics[width=\linewidth]{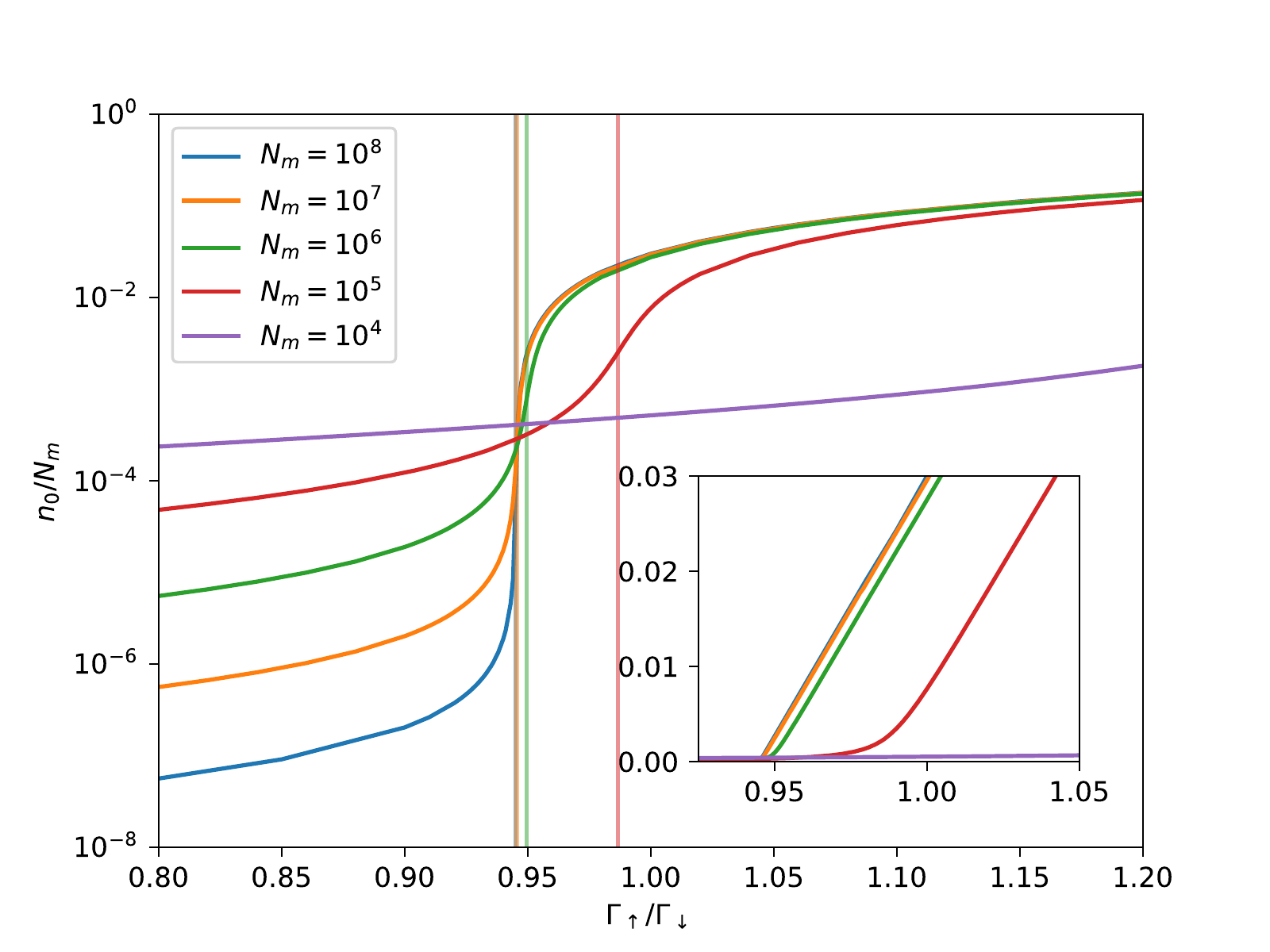}
    \caption{The occupation of the lowest photon mode, scaled by number of molecules, as a function of pump strength. Different lines correspond to different values of $\Nm$. The threshold --- identified as the point of maximum gradient --- has been marked with a vertical line of the same hue. The inset shows a close up of the threshold region on a linear scale. Parameters used: $\Omega_R = 0.1$ eV, $\omega_0=\epsilon$.}
    \label{fig:inout}
\end{figure}

Figure~\ref{fig:inout} shows the occupation of the $k=0$ mode vs pump strength for various different values of $\Nm$.  Threshold can be defined by the steepest gradient on the logarithmic plot.  By extracting this, we see that threshold moves to larger pump power with reducing $\Nm$.  This is particularly pronounced for $\Nm=10^5$, but can also be seen comparing $\Nm=10^6$ to other values.  (The threshold for $\Nm=10^4$ is not seen on this range.)  The increase of threshold when fluctuation effects are larger is consistent with results known in thermal equilibrium~\cite{Nozieres1985,randeria,Keeling2004,Keeling2005}, where the size of changes to the critical density depend on the dimensionless mass parameter $m^\ast = {\Nph}/{\Nm}$ --- the results here are consistent with that trend.

\begin{figure}[htpb]
    \centering
    \includegraphics[width=\linewidth]{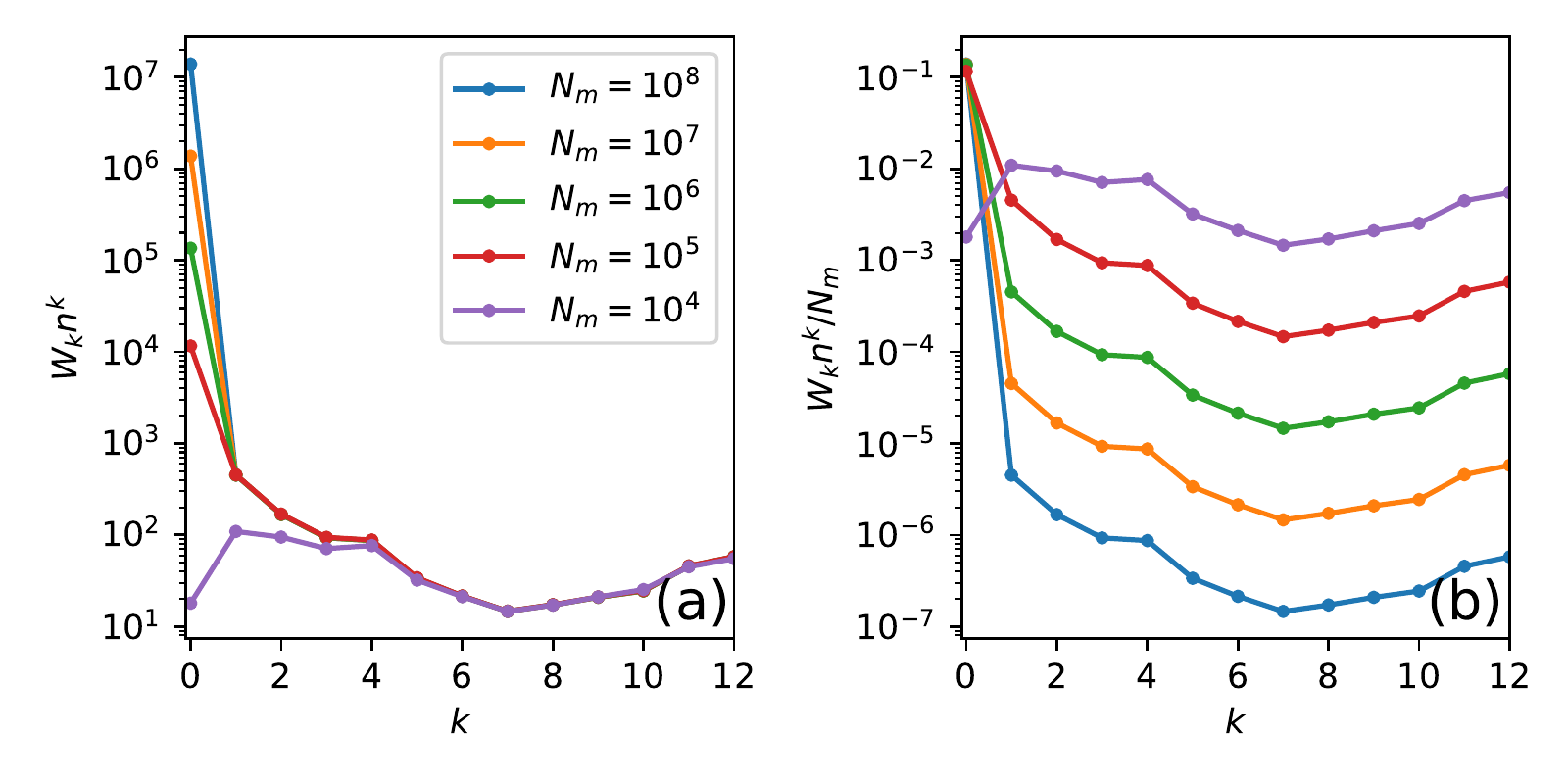}
    \caption{Occupation of the lowest few photon modes at  $\Gamma_\uparrow = 1.2\Gamma_\downarrow$. Panel (a) unscaled, while panel (b) is rescaled by the number of molecules.    Parameters as in Fig.~\ref{fig:inout}.}
    \label{fig:Nm_scaling}
\end{figure}

We also see in Fig.~\ref{fig:inout} that the mode populations scale differently with $\Nm$ above and below threshold.
Above threshold, the population of the lasing mode scales with $\Nm$, so with the rescaling by $\Nm$, the lines collapse to a single curve.  Below threshold, the population is non-extensive (and the curves would match without the rescaling).  This different $\Nm$ scaling provides a useful distinction between truly macroscopic and large but non-extensive population when multimode lasing is seen.  This is illustrated further in Fig.~\ref{fig:Nm_scaling} which shows the mode populations vs $k$ both with and without rescaling. We see a distinction between the lasing mode ($k=0$ in most cases) which scales with $\Nm$ (and shows data collapse at large enough $\Nm$ when rescaled), and fluctuation ($k>0$) which do not depend on $\Nm$.

\section{Large \texorpdfstring{$k$}{k} lasing}
The bright yellow region in the upper right corner of Fig. \ref{fig:phasediagrams}(d) shows abrupt and apparently irregular switch between lasing at $k=0$ and lasing at a large value of $k$. This section provides more details on the nature of this switch and the origin of the irregularities.
\begin{figure}[htpb]
    \centering
    \includegraphics[width=\linewidth]{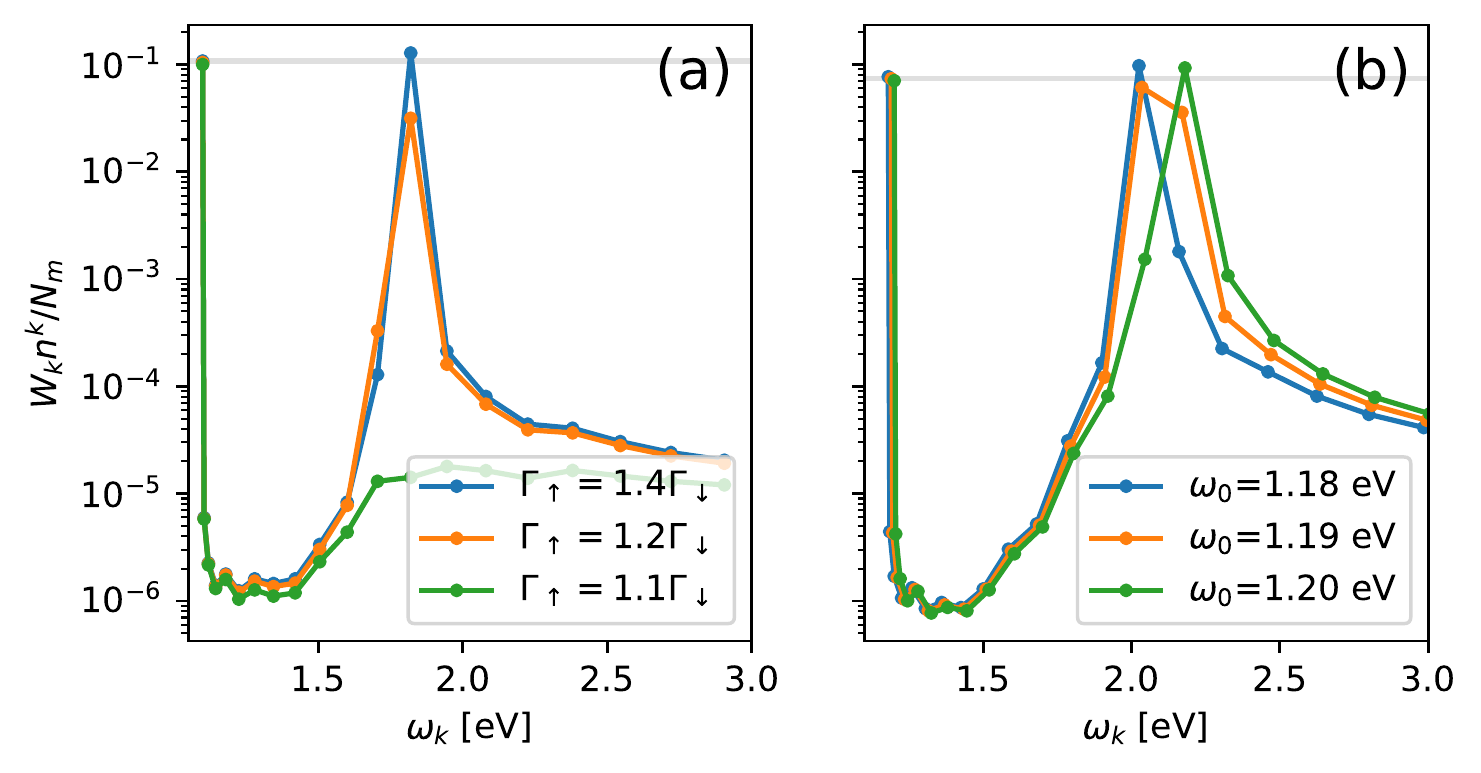}
    \caption{Occupation of photon modes as a function of the bare photon frequency of each mode. The gray horizontal lines mark the occupation of the $k=0$ mode, for comparison to occupation at high $k$ values. (a) Shows behavior at different pumping strengths for a fixed photon dispersion with $\omega_0=1.1$eV.
    (b) Shows how changing $\omega_0$ slightly can result in a sudden change of $k$ value of highest occupied mode when two high $k$ modes have similar occupation.}
    \label{fig:high_k_lasing}
\end{figure}

Figure \ref{fig:high_k_lasing} shows the occupation of photon modes as a function of the bare photon mode frequencies $\omega_k$ for a given value of the pump strength $\Gamma_\uparrow$ and the bare energy of the lowest photon mode $\omega_0$. Panel (a)  shows three different behaviors for different pumping strengths at a fixed $\omega0=1.1$eV. The lowest pumping strength  $\Gamma_\uparrow=1.1\Gamma_\downarrow$ (green line) shows lasing in only one mode: $k=0$. 
When the pump is increased to $\Gamma_\uparrow=1.2\Gamma_\downarrow$ (orange line) there is clearly large occupation at a high $k$ value, but the highest occupation is still in $k=0$ which leads to a dark (blue) square in Fig. \ref{fig:phasediagrams}(d). When the pump gets even larger, like in the topmost line ($\Gamma_\uparrow=1.4\Gamma_\downarrow$, blue line) the occupation of the high $k$ mode overtakes the $k=0$ mode, which results in a bright (yellow) square in the phase diagram.

The irregularities of the boundary in the bright (yellow) region in Fig.~\ref{fig:phasediagrams}(d) can be explained by seeing how this behavior changes with $\omega_0$. Figure \ref{fig:high_k_lasing}(b) also shows photon occupation as a function of the bare mode frequencies but this time the pump strength has been fixed at $\Gamma_\uparrow=1.35\Gamma_\downarrow$ while $\omega_0$ is varied slightly. We can see that both the blue and green line have most occupation at high $k$'s, $k=13$ and $k=14$ respectively, while the orange line has high occupation in both modes. When the occupation is spread over more modes, the $k=0$ becomes the highest occupied mode.

\section{Photoluminescence spectrum}
This section gives more details on calculating the PL spectrum, defined by Eq.~\eqref{eq:PL}.
Two time correlations can be calculated using the quantum regression theorem~\cite{Breuer2002}.  Starting from a steady state $\rho_{ss}$, the two time correlator can then be written
\begin{align*}
\langle a^\dagger_k (t) a_k(0)\rangle = \text{Tr}\left[a^\dagger_k e^{t\mathcal{L}}a_k \rho_{ss}\right],
\end{align*}
where $\mathcal{L}$ is the superoperator defined by $\partial_t \rho = \mathcal{L}\rho$ and Eq.~\eqref{eq:Lindblad}. The equations of motion for the two time correlator can then be calculated by considering the time evolution of the effective density matrix $\tilde\rho(t) = e^{t\mathcal{L}} a_k\rho_{ss}$. The two time correlator is then $\text{Tr}\left[a^\dagger_k \tilde\rho(t)\right]$. Defining $c_i^k(t) = \frac{1}{\Nm}\sum_n e^{-2i\pi k r_n/L} \text{Tr}\left[\lambda_i^{(n)}\tilde\rho(t)\right]$ we can write down a closed set of equations of motion (using the same cumulant approximation as before),
\begin{align}
\partial_t \langle a^\dagger_k (t) a_k(0)\rangle &=
	\left(i\omega_k - \frac{\kappa}{2}\right) \langle a^\dagger_k (t) a_k(0)\rangle + i\Nm B_i^* c_i^k(t),\label{Eq:Fluora}
\\
\partial_t c_i^k(t) &= \xi_{ij}c_j^k(t) + 2f_{ijp} B_j \ell_p\langle a^\dagger_k (t) a_k(0)\rangle.
\label{Eq:Fluorb}
\end{align}
Here we have taken into account that $\ell_p$ is constant in the steady state. As these equations are linear, we can write them on the form $\partial_t \mathbf{C_k} = \mathcal{M}\mathbf{C_k}$ where the vector $\mathbf{C_k} = [\langle a^\dagger_k (t) a_k(0)\rangle, \{c_i^k(t)\}_i]$ and the matrix $\mathcal{M}$ can be read off from Eqs.~(\ref{Eq:Fluora},\ref{Eq:Fluorb}). We can then perform the Fourier transform analytically, assuming there are no eigenvalues with positive real part (which would mean gain in the corresponding mode, which does not happen in a steady state).  We then have the simple form:
\begin{align}
S_k(\nu) &= \int_{-\infty}^{\infty} e^{i\nu t + \mathcal{M}t} \mathbf{C}(0) dt
\nonumber\\&= (i\nu + \mathcal{M})^{-1} \mathbf{C}(0)
= \sum_i \frac{\alpha_i}{\mu_i+i\nu} \ket{r_i},
\end{align}
where $\mu_i$ is the eigenvalue of $\mathcal{M}$ corresponding to the right eigenvector $\ket{r_i}$ and left eigenvector $\bra{l_i}$.  The coefficient $\alpha_i$ is then given by $\alpha_i=\bra{l_i}\mathbf{C}(0)\rangle/\bra{l_i}r_i\rangle$


\begin{thebibliography}{45}%
\makeatletter
\providecommand \@ifxundefined [1]{%
 \@ifx{#1\undefined}
}%
\providecommand \@ifnum [1]{%
 \ifnum #1\expandafter \@firstoftwo
 \else \expandafter \@secondoftwo
 \fi
}%
\providecommand \@ifx [1]{%
 \ifx #1\expandafter \@firstoftwo
 \else \expandafter \@secondoftwo
 \fi
}%
\providecommand \natexlab [1]{#1}%
\providecommand \enquote  [1]{``#1''}%
\providecommand \bibnamefont  [1]{#1}%
\providecommand \bibfnamefont [1]{#1}%
\providecommand \citenamefont [1]{#1}%
\providecommand \href@noop [0]{\@secondoftwo}%
\providecommand \href [0]{\begingroup \@sanitize@url \@href}%
\providecommand \@href[1]{\@@startlink{#1}\@@href}%
\providecommand \@@href[1]{\endgroup#1\@@endlink}%
\providecommand \@sanitize@url [0]{\catcode `\\12\catcode `\$12\catcode
  `\&12\catcode `\#12\catcode `\^12\catcode `\_12\catcode `\%12\relax}%
\providecommand \@@startlink[1]{}%
\providecommand \@@endlink[0]{}%
\providecommand \url  [0]{\begingroup\@sanitize@url \@url }%
\providecommand \@url [1]{\endgroup\@href {#1}{\urlprefix }}%
\providecommand \urlprefix  [0]{URL }%
\providecommand \Eprint [0]{\href }%
\providecommand \doibase [0]{http://dx.doi.org/}%
\providecommand \selectlanguage [0]{\@gobble}%
\providecommand \bibinfo  [0]{\@secondoftwo}%
\providecommand \bibfield  [0]{\@secondoftwo}%
\providecommand \translation [1]{[#1]}%
\providecommand \BibitemOpen [0]{}%
\providecommand \bibitemStop [0]{}%
\providecommand \bibitemNoStop [0]{.\EOS\space}%
\providecommand \EOS [0]{\spacefactor3000\relax}%
\providecommand \BibitemShut  [1]{\csname bibitem#1\endcsname}%
\let\auto@bib@innerbib\@empty
%</preamble>
\bibitem [{\citenamefont {Agranovich}(2009)}]{Agranovich2009a}%
  \BibitemOpen
  \bibfield  {author} {\bibinfo {author} {\bibfnamefont {V.~M.}\ \bibnamefont
  {Agranovich}},\ }\href@noop {} {\emph {\bibinfo {title} {{Excitations in
  Organic Solids}}}}\ (\bibinfo  {publisher} {Oxford University Press},\
  \bibinfo {address} {Oxford},\ \bibinfo {year} {2009})\BibitemShut {NoStop}%
\bibitem [{\citenamefont {Carusotto}\ and\ \citenamefont
  {Ciuti}(2013)}]{Carusotto2013a}%
  \BibitemOpen
  \bibfield  {author} {\bibinfo {author} {\bibfnamefont {I.}~\bibnamefont
  {Carusotto}}\ and\ \bibinfo {author} {\bibfnamefont {C.}~\bibnamefont
  {Ciuti}},\ }\bibfield  {title} {\bibinfo {title} {\emph {{Quantum fluids of
  light}}},\ }\href {\doibase 10.1103/RevModPhys.85.299} {\bibfield  {journal}
  {\bibinfo  {journal} {Rev. Mod. Phys.}\ }\textbf {\bibinfo {volume} {85}},\
  \bibinfo {pages} {299} (\bibinfo {year} {2013})}\BibitemShut {NoStop}%
\bibitem [{\citenamefont {Sanvitto}\ and\ \citenamefont
  {K{\'{e}}na-Cohen}(2016)}]{Sanvitto2016}%
  \BibitemOpen
  \bibfield  {author} {\bibinfo {author} {\bibfnamefont {D.}~\bibnamefont
  {Sanvitto}}\ and\ \bibinfo {author} {\bibfnamefont {S.}~\bibnamefont
  {K{\'{e}}na-Cohen}},\ }\bibfield  {title} {\bibinfo {title} {\emph {{The road
  towards polaritonic devices}}},\ }\href {\doibase 10.1038/nmat4668}
  {\bibfield  {journal} {\bibinfo  {journal} {Nat. Mater.}\ }\textbf {\bibinfo
  {volume} {15}},\ \bibinfo {pages} {1061} (\bibinfo {year}
  {2016})}\BibitemShut {NoStop}%
\bibitem [{\citenamefont {K{\'e}na-Cohen}\ and\ \citenamefont
  {Forrest}(2010)}]{kena2010room}%
  \BibitemOpen
  \bibfield  {author} {\bibinfo {author} {\bibfnamefont {S.}~\bibnamefont
  {K{\'e}na-Cohen}}\ and\ \bibinfo {author} {\bibfnamefont {S.}~\bibnamefont
  {Forrest}},\ }\bibfield  {title} {\bibinfo {title} {\emph {Room-temperature
  polariton lasing in an organic single-crystal microcavity}},\ }\href
  {\doibase https://doi.org/10.1038/nphoton.2010.86} {\bibfield  {journal}
  {\bibinfo  {journal} {Nat. Photon.}\ }\textbf {\bibinfo {volume} {4}},\
  \bibinfo {pages} {371} (\bibinfo {year} {2010})}\BibitemShut {NoStop}%
\bibitem [{\citenamefont {Plumhof}\ \emph {et~al.}(2014)\citenamefont
  {Plumhof}, \citenamefont {St{\"{o}}ferle}, \citenamefont {Mai}, \citenamefont
  {Scherf},\ and\ \citenamefont {Mahrt}}]{Plumhof14}%
  \BibitemOpen
  \bibfield  {author} {\bibinfo {author} {\bibfnamefont {J.~D.}\ \bibnamefont
  {Plumhof}}, \bibinfo {author} {\bibfnamefont {T.}~\bibnamefont
  {St{\"{o}}ferle}}, \bibinfo {author} {\bibfnamefont {L.}~\bibnamefont {Mai}},
  \bibinfo {author} {\bibfnamefont {U.}~\bibnamefont {Scherf}}, \ and\ \bibinfo
  {author} {\bibfnamefont {R.~F.}\ \bibnamefont {Mahrt}},\ }\bibfield  {title}
  {\bibinfo {title} {\emph {{Room-temperature Bose-Einstein condensation of
  cavity exciton-polaritons in a polymer.}}},\ }\href {\doibase
  10.1038/nmat3825} {\bibfield  {journal} {\bibinfo  {journal} {Nat. Mater.}\
  }\textbf {\bibinfo {volume} {13}},\ \bibinfo {pages} {247} (\bibinfo {year}
  {2014})}\BibitemShut {NoStop}%
\bibitem [{\citenamefont {Daskalakis}\ \emph {et~al.}(2014)\citenamefont
  {Daskalakis}, \citenamefont {Maier}, \citenamefont {Murray},\ and\
  \citenamefont {K{\'{e}}na-Cohen}}]{Daskalakis2014}%
  \BibitemOpen
  \bibfield  {author} {\bibinfo {author} {\bibfnamefont {K.~S.}\ \bibnamefont
  {Daskalakis}}, \bibinfo {author} {\bibfnamefont {S.~A.}\ \bibnamefont
  {Maier}}, \bibinfo {author} {\bibfnamefont {R.}~\bibnamefont {Murray}}, \
  and\ \bibinfo {author} {\bibfnamefont {S.}~\bibnamefont {K{\'{e}}na-Cohen}},\
  }\bibfield  {title} {\bibinfo {title} {\emph {{Nonlinear interactions in an
  organic polariton condensate.}}},\ }\href {\doibase 10.1038/nmat3874}
  {\bibfield  {journal} {\bibinfo  {journal} {Nat. Mater.}\ }\textbf {\bibinfo
  {volume} {13}},\ \bibinfo {pages} {271} (\bibinfo {year} {2014})}\BibitemShut
  {NoStop}%
\bibitem [{\citenamefont {Dietrich}\ \emph {et~al.}(2016)\citenamefont
  {Dietrich}, \citenamefont {Steude}, \citenamefont {Tropf}, \citenamefont
  {Schubert}, \citenamefont {Kronenberg}, \citenamefont {Ostermann},
  \citenamefont {H{\"{o}}fling},\ and\ \citenamefont {Gather}}]{dietrich16}%
  \BibitemOpen
  \bibfield  {author} {\bibinfo {author} {\bibfnamefont {C.~P.}\ \bibnamefont
  {Dietrich}}, \bibinfo {author} {\bibfnamefont {A.}~\bibnamefont {Steude}},
  \bibinfo {author} {\bibfnamefont {L.}~\bibnamefont {Tropf}}, \bibinfo
  {author} {\bibfnamefont {M.}~\bibnamefont {Schubert}}, \bibinfo {author}
  {\bibfnamefont {N.~M.}\ \bibnamefont {Kronenberg}}, \bibinfo {author}
  {\bibfnamefont {K.}~\bibnamefont {Ostermann}}, \bibinfo {author}
  {\bibfnamefont {S.}~\bibnamefont {H{\"{o}}fling}}, \ and\ \bibinfo {author}
  {\bibfnamefont {M.~C.}\ \bibnamefont {Gather}},\ }\bibfield  {title}
  {\bibinfo {title} {\emph {{An exciton-polariton laser based on biologically
  produced fluorescent protein}}},\ }\href {\doibase 10.1126/sciadv.1600666}
  {\bibfield  {journal} {\bibinfo  {journal} {Sci. Adv.}\ }\textbf {\bibinfo
  {volume} {2}},\ \bibinfo {pages} {e1600666} (\bibinfo {year}
  {2016})}\BibitemShut {NoStop}%
\bibitem [{\citenamefont {Cookson}\ \emph {et~al.}(2017)\citenamefont
  {Cookson}, \citenamefont {Georgiou}, \citenamefont {Zasedatelev},
  \citenamefont {Grant}, \citenamefont {Virgili}, \citenamefont {Cavazzini},
  \citenamefont {Galeotti}, \citenamefont {Clark}, \citenamefont {Berloff},
  \citenamefont {Lidzey},\ and\ \citenamefont {Lagoudakis}}]{cookson2017}%
  \BibitemOpen
  \bibfield  {author} {\bibinfo {author} {\bibfnamefont {T.}~\bibnamefont
  {Cookson}}, \bibinfo {author} {\bibfnamefont {K.}~\bibnamefont {Georgiou}},
  \bibinfo {author} {\bibfnamefont {A.}~\bibnamefont {Zasedatelev}}, \bibinfo
  {author} {\bibfnamefont {R.~T.}\ \bibnamefont {Grant}}, \bibinfo {author}
  {\bibfnamefont {T.}~\bibnamefont {Virgili}}, \bibinfo {author} {\bibfnamefont
  {M.}~\bibnamefont {Cavazzini}}, \bibinfo {author} {\bibfnamefont
  {F.}~\bibnamefont {Galeotti}}, \bibinfo {author} {\bibfnamefont
  {C.}~\bibnamefont {Clark}}, \bibinfo {author} {\bibfnamefont {N.~G.}\
  \bibnamefont {Berloff}}, \bibinfo {author} {\bibfnamefont {D.~G.}\
  \bibnamefont {Lidzey}}, \ and\ \bibinfo {author} {\bibfnamefont {P.~G.}\
  \bibnamefont {Lagoudakis}},\ }\bibfield  {title} {\bibinfo {title} {\emph {A
  Yellow Polariton Condensate in a Dye Filled Microcavity}},\ }\href {\doibase
  10.1002/adom.201700203} {\bibfield  {journal} {\bibinfo  {journal} {Adv. Opt.
  Mater.}\ }\textbf {\bibinfo {volume} {5}},\ \bibinfo {pages} {1700203}
  (\bibinfo {year} {2017})}\BibitemShut {NoStop}%
\bibitem [{\citenamefont {Lerario}\ \emph {et~al.}(2017)\citenamefont
  {Lerario}, \citenamefont {Fieramosca}, \citenamefont {Barachati},
  \citenamefont {Ballarini}, \citenamefont {Daskalakis}, \citenamefont
  {Dominici}, \citenamefont {De~Giorgi}, \citenamefont {Maier}, \citenamefont
  {Gigli}, \citenamefont {Kéna-Cohen},\ and\ \citenamefont
  {Sanvitto}}]{Lerario2017}%
  \BibitemOpen
  \bibfield  {author} {\bibinfo {author} {\bibfnamefont {G.}~\bibnamefont
  {Lerario}}, \bibinfo {author} {\bibfnamefont {A.}~\bibnamefont {Fieramosca}},
  \bibinfo {author} {\bibfnamefont {F.}~\bibnamefont {Barachati}}, \bibinfo
  {author} {\bibfnamefont {D.}~\bibnamefont {Ballarini}}, \bibinfo {author}
  {\bibfnamefont {K.~S.}\ \bibnamefont {Daskalakis}}, \bibinfo {author}
  {\bibfnamefont {L.}~\bibnamefont {Dominici}}, \bibinfo {author}
  {\bibfnamefont {M.}~\bibnamefont {De~Giorgi}}, \bibinfo {author}
  {\bibfnamefont {S.~A.}\ \bibnamefont {Maier}}, \bibinfo {author}
  {\bibfnamefont {G.}~\bibnamefont {Gigli}}, \bibinfo {author} {\bibfnamefont
  {S.}~\bibnamefont {Kéna-Cohen}}, \ and\ \bibinfo {author} {\bibfnamefont
  {D.}~\bibnamefont {Sanvitto}},\ }\bibfield  {title} {\bibinfo {title} {\emph
  {Room-temperature superfluidity in a polariton condensate}},\ }\href
  {https://doi.org/10.1038/nphys4147} {\bibfield  {journal} {\bibinfo
  {journal} {Nat. Phys.}\ }\textbf {\bibinfo {volume} {13}},\ \bibinfo {pages}
  {837} (\bibinfo {year} {2017})}\BibitemShut {NoStop}%
\bibitem [{\citenamefont {Ramezani}\ \emph {et~al.}(2017)\citenamefont
  {Ramezani}, \citenamefont {Halpin}, \citenamefont
  {Fern\'{a}ndez-Dom\'{i}nguez}, \citenamefont {Feist}, \citenamefont
  {Rodriguez}, \citenamefont {Garcia-Vidal},\ and\ \citenamefont
  {Rivas}}]{Ramezani17}%
  \BibitemOpen
  \bibfield  {author} {\bibinfo {author} {\bibfnamefont {M.}~\bibnamefont
  {Ramezani}}, \bibinfo {author} {\bibfnamefont {A.}~\bibnamefont {Halpin}},
  \bibinfo {author} {\bibfnamefont {A.~I.}\ \bibnamefont
  {Fern\'{a}ndez-Dom\'{i}nguez}}, \bibinfo {author} {\bibfnamefont
  {J.}~\bibnamefont {Feist}}, \bibinfo {author} {\bibfnamefont {S.~R.-K.}\
  \bibnamefont {Rodriguez}}, \bibinfo {author} {\bibfnamefont {F.~J.}\
  \bibnamefont {Garcia-Vidal}}, \ and\ \bibinfo {author} {\bibfnamefont
  {J.~G.}\ \bibnamefont {Rivas}},\ }\bibfield  {title} {\bibinfo {title} {\emph
  {Plasmon-exciton-polariton lasing}},\ }\href {\doibase
  10.1364/OPTICA.4.000031} {\bibfield  {journal} {\bibinfo  {journal} {Optica}\
  }\textbf {\bibinfo {volume} {4}},\ \bibinfo {pages} {31} (\bibinfo {year}
  {2017})}\BibitemShut {NoStop}%
\bibitem [{\citenamefont {Scafirimuto}\ \emph {et~al.}(2018)\citenamefont
  {Scafirimuto}, \citenamefont {Urbonas}, \citenamefont {Scherf}, \citenamefont
  {Mahrt},\ and\ \citenamefont {St\"oferle}}]{scafirimuto2018}%
  \BibitemOpen
  \bibfield  {author} {\bibinfo {author} {\bibfnamefont {F.}~\bibnamefont
  {Scafirimuto}}, \bibinfo {author} {\bibfnamefont {D.}~\bibnamefont
  {Urbonas}}, \bibinfo {author} {\bibfnamefont {U.}~\bibnamefont {Scherf}},
  \bibinfo {author} {\bibfnamefont {R.~F.}\ \bibnamefont {Mahrt}}, \ and\
  \bibinfo {author} {\bibfnamefont {T.}~\bibnamefont {St\"oferle}},\ }\bibfield
   {title} {\bibinfo {title} {\emph {Room-Temperature Exciton-Polariton
  Condensation in a Tunable Zero-Dimensional Microcavity}},\ }\href {\doibase
  10.1021/acsphotonics.7b00557} {\bibfield  {journal} {\bibinfo  {journal} {ACS
  Photonics}\ }\textbf {\bibinfo {volume} {5}},\ \bibinfo {pages} {85}
  (\bibinfo {year} {2018})}\BibitemShut {NoStop}%
\bibitem [{\citenamefont {Rajendran}\ \emph {et~al.}(2019)\citenamefont
  {Rajendran}, \citenamefont {Wei}, \citenamefont {Ohadi}, \citenamefont
  {Ruseckas}, \citenamefont {Turnbull},\ and\ \citenamefont
  {Samuel}}]{Rajendran2019}%
  \BibitemOpen
  \bibfield  {author} {\bibinfo {author} {\bibfnamefont {S.~K.}\ \bibnamefont
  {Rajendran}}, \bibinfo {author} {\bibfnamefont {M.}~\bibnamefont {Wei}},
  \bibinfo {author} {\bibfnamefont {H.}~\bibnamefont {Ohadi}}, \bibinfo
  {author} {\bibfnamefont {A.}~\bibnamefont {Ruseckas}}, \bibinfo {author}
  {\bibfnamefont {G.~A.}\ \bibnamefont {Turnbull}}, \ and\ \bibinfo {author}
  {\bibfnamefont {I.~D.~W.}\ \bibnamefont {Samuel}},\ }\bibfield  {title}
  {\bibinfo {title} {\emph {Low Threshold Polariton Lasing from a
  Solution-Processed Organic Semiconductor in a Planar Microcavity}},\ }\href
  {\doibase 10.1002/adom.201801791} {\bibfield  {journal} {\bibinfo  {journal}
  {Adv. Opt. Mater.}\ }\textbf {\bibinfo {volume} {7}},\ \bibinfo {pages}
  {1801791} (\bibinfo {year} {2019})}\BibitemShut {NoStop}%
\bibitem [{\citenamefont {Wei}\ \emph {et~al.}(2019)\citenamefont {Wei},
  \citenamefont {Rajendran}, \citenamefont {Ohadi}, \citenamefont {Tropf},
  \citenamefont {Gather}, \citenamefont {Turnbull},\ and\ \citenamefont
  {Samuel.}}]{Wei2019}%
  \BibitemOpen
  \bibfield  {author} {\bibinfo {author} {\bibfnamefont {M.}~\bibnamefont
  {Wei}}, \bibinfo {author} {\bibfnamefont {S.~K.}\ \bibnamefont {Rajendran}},
  \bibinfo {author} {\bibfnamefont {H.}~\bibnamefont {Ohadi}}, \bibinfo
  {author} {\bibfnamefont {L.}~\bibnamefont {Tropf}}, \bibinfo {author}
  {\bibfnamefont {M.~C.}\ \bibnamefont {Gather}}, \bibinfo {author}
  {\bibfnamefont {G.~A.}\ \bibnamefont {Turnbull}}, \ and\ \bibinfo {author}
  {\bibfnamefont {I.~D.~W.}\ \bibnamefont {Samuel.}},\ }\bibfield  {title}
  {\bibinfo {title} {\emph {Low threshold polariton lasing in a highly
  disordered conjugated polymer}},\ }\href {\doibase 10.1364/OPTICA.6.001124}
  {\bibfield  {journal} {\bibinfo  {journal} {Optica}\ }\textbf {\bibinfo
  {volume} {6}},\ \bibinfo {pages} {1124} (\bibinfo {year} {2019})}\BibitemShut
  {NoStop}%
\bibitem [{\citenamefont {V\"akev\"ainen}\ \emph {et~al.}(2019)\citenamefont
  {V\"akev\"ainen}, \citenamefont {Moilanen}, \citenamefont {Ne\v{c}ada},
  \citenamefont {Hakala}, \citenamefont {Daskalakis},\ and\ \citenamefont
  {T\"orm\"a}}]{Vakevainen2019}%
  \BibitemOpen
  \bibfield  {author} {\bibinfo {author} {\bibfnamefont {A.~I.}\ \bibnamefont
  {V\"akev\"ainen}}, \bibinfo {author} {\bibfnamefont {A.~J.}\ \bibnamefont
  {Moilanen}}, \bibinfo {author} {\bibfnamefont {M.}~\bibnamefont
  {Ne\v{c}ada}}, \bibinfo {author} {\bibfnamefont {T.~K.}\ \bibnamefont
  {Hakala}}, \bibinfo {author} {\bibfnamefont {K.~S.}\ \bibnamefont
  {Daskalakis}}, \ and\ \bibinfo {author} {\bibfnamefont {P.}~\bibnamefont
  {T\"orm\"a}},\ }\href@noop {} {\bibinfo {title} {\emph {Sub-picosecond
  thermalization dynamics in condensation of strongly coupled lattice
  plasmons}}} (\bibinfo {year} {2019}),\ \bibinfo {note} {preprint},\ \Eprint
  {http://arxiv.org/abs/1905.07609} {1905.07609} \BibitemShut {NoStop}%
\bibitem [{\citenamefont {Yagafarov}\ \emph {et~al.}(2020)\citenamefont
  {Yagafarov}, \citenamefont {Sannikov}, \citenamefont {Zasedatelev},
  \citenamefont {Georgiou}, \citenamefont {Baranikov}, \citenamefont
  {Kyriienko}, \citenamefont {Shelykh}, \citenamefont {Gai}, \citenamefont
  {Shen}, \citenamefont {Lidzey},\ and\ \citenamefont
  {Lagoudakis}}]{yagafarov2020blueshift}%
  \BibitemOpen
  \bibfield  {author} {\bibinfo {author} {\bibfnamefont {T.}~\bibnamefont
  {Yagafarov}}, \bibinfo {author} {\bibfnamefont {D.}~\bibnamefont {Sannikov}},
  \bibinfo {author} {\bibfnamefont {A.}~\bibnamefont {Zasedatelev}}, \bibinfo
  {author} {\bibfnamefont {K.}~\bibnamefont {Georgiou}}, \bibinfo {author}
  {\bibfnamefont {A.}~\bibnamefont {Baranikov}}, \bibinfo {author}
  {\bibfnamefont {O.}~\bibnamefont {Kyriienko}}, \bibinfo {author}
  {\bibfnamefont {I.}~\bibnamefont {Shelykh}}, \bibinfo {author} {\bibfnamefont
  {L.}~\bibnamefont {Gai}}, \bibinfo {author} {\bibfnamefont {Z.}~\bibnamefont
  {Shen}}, \bibinfo {author} {\bibfnamefont {D.~G.}\ \bibnamefont {Lidzey}}, \
  and\ \bibinfo {author} {\bibfnamefont {P.}~\bibnamefont {Lagoudakis}},\
  }\bibfield  {title} {\bibinfo {title} {\emph {Mechanisms of blueshifts in
  organic polariton condensates}},\ }\href {\doibase 10.1038/s42005-019-0278-6}
  {\bibfield  {journal} {\bibinfo  {journal} {Commun Phys}\ }\textbf {\bibinfo
  {volume} {3}},\ \bibinfo {pages} {18} (\bibinfo {year} {2020})}\BibitemShut
  {NoStop}%
\bibitem [{\citenamefont {Keeling}\ and\ \citenamefont
  {Kéna-Cohen}(2020)}]{keeling2020review}%
  \BibitemOpen
  \bibfield  {author} {\bibinfo {author} {\bibfnamefont {J.}~\bibnamefont
  {Keeling}}\ and\ \bibinfo {author} {\bibfnamefont {S.}~\bibnamefont
  {Kéna-Cohen}},\ }\bibfield  {title} {\bibinfo {title} {\emph
  {Bose–Einstein Condensation of Exciton-Polaritons in Organic
  Microcavities}},\ }\href {\doibase 10.1146/annurev-physchem-010920-102509}
  {\bibfield  {journal} {\bibinfo  {journal} {Ann. Rev. Phys. Chem.}\ }\textbf
  {\bibinfo {volume} {71}},\ \bibinfo {pages} {null} (\bibinfo {year}
  {2020})}\BibitemShut {NoStop}%
\bibitem [{\citenamefont {Michetti}\ \emph {et~al.}(2009)\citenamefont
  {Michetti}, \citenamefont {{La Rocca}},\ and\ \citenamefont
  {Rocca}}]{Michetti2009}%
  \BibitemOpen
  \bibfield  {author} {\bibinfo {author} {\bibfnamefont {P.}~\bibnamefont
  {Michetti}}, \bibinfo {author} {\bibfnamefont {G.}~\bibnamefont {{La
  Rocca}}}, \ and\ \bibinfo {author} {\bibfnamefont {G.~C.~L.}\ \bibnamefont
  {Rocca}},\ }\bibfield  {title} {\bibinfo {title} {\emph {{Exciton-phonon
  scattering and photoexcitation dynamics in J-aggregate microcavities}}},\
  }\href {\doibase 10.1103/PhysRevB.79.035325} {\bibfield  {journal} {\bibinfo
  {journal} {Phys. Rev. B}\ }\textbf {\bibinfo {volume} {79}},\ \bibinfo
  {pages} {035325} (\bibinfo {year} {2009})}\BibitemShut {NoStop}%
\bibitem [{\citenamefont {Fontanesi}\ and\ \citenamefont {{La
  Rocca}}(2009)}]{Fontanesi2009}%
  \BibitemOpen
  \bibfield  {author} {\bibinfo {author} {\bibfnamefont {L.}~\bibnamefont
  {Fontanesi}}\ and\ \bibinfo {author} {\bibfnamefont {G.~C.}\ \bibnamefont
  {{La Rocca}}},\ }\bibfield  {title} {\bibinfo {title} {\emph {{Organic-based
  microcavities with vibronic progressions: Linear spectroscopy}}},\ }\href
  {\doibase 10.1103/PhysRevB.80.235313} {\bibfield  {journal} {\bibinfo
  {journal} {Phys. Rev. B}\ }\textbf {\bibinfo {volume} {80}},\ \bibinfo
  {pages} {235313} (\bibinfo {year} {2009})}\BibitemShut {NoStop}%
\bibitem [{\citenamefont {Mazza}\ and\ \citenamefont {{La
  Rocca}}(2009)}]{Mazza2009}%
  \BibitemOpen
  \bibfield  {author} {\bibinfo {author} {\bibfnamefont {L.}~\bibnamefont
  {Mazza}}\ and\ \bibinfo {author} {\bibfnamefont {G.~C.}\ \bibnamefont {{La
  Rocca}}},\ }\bibfield  {title} {\bibinfo {title} {\emph {{Organic-based
  microcavities with vibronic progressions: Photoluminescence}}},\ }\href
  {\doibase 10.1103/PhysRevB.80.235314} {\bibfield  {journal} {\bibinfo
  {journal} {Phys. Rev. B}\ }\textbf {\bibinfo {volume} {80}},\ \bibinfo
  {pages} {235314} (\bibinfo {year} {2009})}\BibitemShut {NoStop}%
\bibitem [{\citenamefont {Mazza}\ \emph {et~al.}(2013)\citenamefont {Mazza},
  \citenamefont {K\'ena-Cohen}, \citenamefont {Michetti},\ and\ \citenamefont
  {La~Rocca}}]{mazza2013microscopic}%
  \BibitemOpen
  \bibfield  {author} {\bibinfo {author} {\bibfnamefont {L.}~\bibnamefont
  {Mazza}}, \bibinfo {author} {\bibfnamefont {S.}~\bibnamefont {K\'ena-Cohen}},
  \bibinfo {author} {\bibfnamefont {P.}~\bibnamefont {Michetti}}, \ and\
  \bibinfo {author} {\bibfnamefont {G.~C.}\ \bibnamefont {La~Rocca}},\
  }\bibfield  {title} {\bibinfo {title} {\emph {Microscopic theory of polariton
  lasing via vibronically assisted scattering}},\ }\href {\doibase
  10.1103/PhysRevB.88.075321} {\bibfield  {journal} {\bibinfo  {journal} {Phys.
  Rev. B}\ }\textbf {\bibinfo {volume} {88}},\ \bibinfo {pages} {075321}
  (\bibinfo {year} {2013})}\BibitemShut {NoStop}%
\bibitem [{\citenamefont {Coles}\ \emph {et~al.}(2011)\citenamefont {Coles},
  \citenamefont {Michetti}, \citenamefont {Clark}, \citenamefont {Tsoi},
  \citenamefont {Adawi}, \citenamefont {Kim},\ and\ \citenamefont
  {Lidzey}}]{coles2011vibrationally}%
  \BibitemOpen
  \bibfield  {author} {\bibinfo {author} {\bibfnamefont {D.~M.}\ \bibnamefont
  {Coles}}, \bibinfo {author} {\bibfnamefont {P.}~\bibnamefont {Michetti}},
  \bibinfo {author} {\bibfnamefont {C.}~\bibnamefont {Clark}}, \bibinfo
  {author} {\bibfnamefont {W.~C.}\ \bibnamefont {Tsoi}}, \bibinfo {author}
  {\bibfnamefont {A.~M.}\ \bibnamefont {Adawi}}, \bibinfo {author}
  {\bibfnamefont {J.-S.}\ \bibnamefont {Kim}}, \ and\ \bibinfo {author}
  {\bibfnamefont {D.~G.}\ \bibnamefont {Lidzey}},\ }\bibfield  {title}
  {\bibinfo {title} {\emph {Vibrationally Assisted Polariton-Relaxation
  Processes in Strongly Coupled Organic-Semiconductor Microcavities}},\ }\href
  {\doibase 10.1002/adfm.201100756} {\bibfield  {journal} {\bibinfo  {journal}
  {Adv. Funct. Mater.}\ }\textbf {\bibinfo {volume} {21}},\ \bibinfo {pages}
  {3691} (\bibinfo {year} {2011})}\BibitemShut {NoStop}%
\bibitem [{\citenamefont {{\'{C}}wik}\ \emph {et~al.}(2014)\citenamefont
  {{\'{C}}wik}, \citenamefont {Reja}, \citenamefont {Littlewood},\ and\
  \citenamefont {Keeling}}]{Cwik2014}%
  \BibitemOpen
  \bibfield  {author} {\bibinfo {author} {\bibfnamefont {J.~A.}\ \bibnamefont
  {{\'{C}}wik}}, \bibinfo {author} {\bibfnamefont {S.}~\bibnamefont {Reja}},
  \bibinfo {author} {\bibfnamefont {P.~B.}\ \bibnamefont {Littlewood}}, \ and\
  \bibinfo {author} {\bibfnamefont {J.}~\bibnamefont {Keeling}},\ }\bibfield
  {title} {\bibinfo {title} {\emph {{Polariton condensation with saturable
  molecules dressed by vibrational modes}}},\ }\href {\doibase
  10.1209/0295-5075/105/47009} {\bibfield  {journal} {\bibinfo  {journal} {Eur.
  Lett.}\ }\textbf {\bibinfo {volume} {105}},\ \bibinfo {pages} {47009}
  (\bibinfo {year} {2014})}\BibitemShut {NoStop}%
\bibitem [{\citenamefont {Galego}\ \emph {et~al.}(2015)\citenamefont {Galego},
  \citenamefont {Garcia-Vidal},\ and\ \citenamefont {Feist}}]{Galego15}%
  \BibitemOpen
  \bibfield  {author} {\bibinfo {author} {\bibfnamefont {J.}~\bibnamefont
  {Galego}}, \bibinfo {author} {\bibfnamefont {F.~J.}\ \bibnamefont
  {Garcia-Vidal}}, \ and\ \bibinfo {author} {\bibfnamefont {J.}~\bibnamefont
  {Feist}},\ }\bibfield  {title} {\bibinfo {title} {\emph {{Cavity-Induced
  Modifications of Molecular Structure in the Strong-Coupling Regime}}},\
  }\href {\doibase 10.1103/PhysRevX.5.041022} {\bibfield  {journal} {\bibinfo
  {journal} {Phys. Rev. X}\ }\textbf {\bibinfo {volume} {5}},\ \bibinfo {pages}
  {41022} (\bibinfo {year} {2015})}\BibitemShut {NoStop}%
\bibitem [{\citenamefont {Herrera}\ and\ \citenamefont
  {Spano}(2017)}]{Herrera2016a}%
  \BibitemOpen
  \bibfield  {author} {\bibinfo {author} {\bibfnamefont {F.}~\bibnamefont
  {Herrera}}\ and\ \bibinfo {author} {\bibfnamefont {F.~C.}\ \bibnamefont
  {Spano}},\ }\bibfield  {title} {\bibinfo {title} {\emph {{Dark Vibronic
  Polaritons and the Spectroscopy of Organic Microcavities}}},\ }\href
  {\doibase 10.1103/PhysRevLett.118.223601} {\bibfield  {journal} {\bibinfo
  {journal} {Phys. Rev. Lett.}\ }\textbf {\bibinfo {volume} {118}},\ \bibinfo
  {pages} {223601} (\bibinfo {year} {2017})}\BibitemShut {NoStop}%
\bibitem [{\citenamefont {del Pino}\ \emph
  {et~al.}(2018{\natexlab{a}})\citenamefont {del Pino}, \citenamefont
  {Schr{\"{o}}der}, \citenamefont {Chin}, \citenamefont {Feist},\ and\
  \citenamefont {Garcia-Vidal}}]{delPino2018:tensorA}%
  \BibitemOpen
  \bibfield  {author} {\bibinfo {author} {\bibfnamefont {J.}~\bibnamefont {del
  Pino}}, \bibinfo {author} {\bibfnamefont {F.~A. Y.~N.}\ \bibnamefont
  {Schr{\"{o}}der}}, \bibinfo {author} {\bibfnamefont {A.~W.}\ \bibnamefont
  {Chin}}, \bibinfo {author} {\bibfnamefont {J.}~\bibnamefont {Feist}}, \ and\
  \bibinfo {author} {\bibfnamefont {F.~J.}\ \bibnamefont {Garcia-Vidal}},\
  }\bibfield  {title} {\bibinfo {title} {\emph {Tensor Network Simulation of
  Non-Markovian Dynamics in Organic Polaritons}},\ }\href {\doibase
  10.1103/PhysRevLett.121.227401} {\bibfield  {journal} {\bibinfo  {journal}
  {Phys. Rev. Lett.}\ }\textbf {\bibinfo {volume} {121}},\ \bibinfo {pages}
  {227401} (\bibinfo {year} {2018}{\natexlab{a}})}\BibitemShut {NoStop}%
\bibitem [{\citenamefont {del Pino}\ \emph
  {et~al.}(2018{\natexlab{b}})\citenamefont {del Pino}, \citenamefont
  {Schr\"oder}, \citenamefont {Chin}, \citenamefont {Feist},\ and\
  \citenamefont {Garcia-Vidal}}]{delPino2018:tensorB}%
  \BibitemOpen
  \bibfield  {author} {\bibinfo {author} {\bibfnamefont {J.}~\bibnamefont {del
  Pino}}, \bibinfo {author} {\bibfnamefont {F.~A. Y.~N.}\ \bibnamefont
  {Schr\"oder}}, \bibinfo {author} {\bibfnamefont {A.~W.}\ \bibnamefont
  {Chin}}, \bibinfo {author} {\bibfnamefont {J.}~\bibnamefont {Feist}}, \ and\
  \bibinfo {author} {\bibfnamefont {F.~J.}\ \bibnamefont {Garcia-Vidal}},\
  }\bibfield  {title} {\bibinfo {title} {\emph {Tensor network simulation of
  polaron-polaritons in organic microcavities}},\ }\href {\doibase
  10.1103/PhysRevB.98.165416} {\bibfield  {journal} {\bibinfo  {journal} {Phys.
  Rev. B}\ }\textbf {\bibinfo {volume} {98}},\ \bibinfo {pages} {165416}
  (\bibinfo {year} {2018}{\natexlab{b}})}\BibitemShut {NoStop}%
\bibitem [{\citenamefont {Haken}(1970)}]{haken70}%
  \BibitemOpen
  \bibfield  {author} {\bibinfo {author} {\bibfnamefont {H.}~\bibnamefont
  {Haken}},\ }\bibfield  {title} {\bibinfo {title} {\emph {{The semiclassical
  and quantum theory of the laser}}},\ }in\ \href@noop {} {\emph {\bibinfo
  {booktitle} {Quantum Optics}}},\ \bibinfo {editor} {edited by\ \bibinfo
  {editor} {\bibfnamefont {S.~M.}\ \bibnamefont {Kay}}\ and\ \bibinfo {editor}
  {\bibfnamefont {A.}~\bibnamefont {Maitland}}}\ (\bibinfo  {publisher}
  {Academic Press},\ \bibinfo {address} {New York},\ \bibinfo {year} {1970})\
  p.\ \bibinfo {pages} {201}\BibitemShut {NoStop}%
\bibitem [{\citenamefont {Scully}\ and\ \citenamefont
  {Zubairy}(1997)}]{Scully1997}%
  \BibitemOpen
  \bibfield  {author} {\bibinfo {author} {\bibfnamefont {M.~O.}\ \bibnamefont
  {Scully}}\ and\ \bibinfo {author} {\bibfnamefont {M.~S.}\ \bibnamefont
  {Zubairy}},\ }\href {\doibase 10.1017/CBO9780511813993} {\emph {\bibinfo
  {title} {{Quantum Optics}}}}\ (\bibinfo  {publisher} {Cambridge Univ.
  Press},\ \bibinfo {address} {Cambridge},\ \bibinfo {year} {1997})\BibitemShut
  {NoStop}%
\bibitem [{\citenamefont {Strashko}\ \emph {et~al.}(2018)\citenamefont
  {Strashko}, \citenamefont {Kirton},\ and\ \citenamefont
  {Keeling}}]{Strashko2018}%
  \BibitemOpen
  \bibfield  {author} {\bibinfo {author} {\bibfnamefont {A.}~\bibnamefont
  {Strashko}}, \bibinfo {author} {\bibfnamefont {P.}~\bibnamefont {Kirton}}, \
  and\ \bibinfo {author} {\bibfnamefont {J.}~\bibnamefont {Keeling}},\
  }\bibfield  {title} {\bibinfo {title} {\emph {{Organic Polariton Lasing and
  the Weak to Strong Coupling Crossover}}},\ }\href {\doibase
  10.1103/PhysRevLett.121.193601} {\bibfield  {journal} {\bibinfo  {journal}
  {Phys. Rev. Lett.}\ }\textbf {\bibinfo {volume} {121}},\ \bibinfo {pages}
  {193601} (\bibinfo {year} {2018})}\BibitemShut {NoStop}%
\bibitem [{\citenamefont {Nozi\`eres}\ and\ \citenamefont
  {Schmitt-Rink}(1985)}]{Nozieres1985}%
  \BibitemOpen
  \bibfield  {author} {\bibinfo {author} {\bibfnamefont {P.}~\bibnamefont
  {Nozi\`eres}}\ and\ \bibinfo {author} {\bibfnamefont {S.}~\bibnamefont
  {Schmitt-Rink}},\ }\bibfield  {title} {\bibinfo {title} {\emph {{Bose
  condensation in an attractive fermion gas: From weak to strong coupling
  superconductivity}}},\ }\href {\doibase 10.1007/BF00683774} {\bibfield
  {journal} {\bibinfo  {journal} {J. Low Temp. Phys.}\ }\textbf {\bibinfo
  {volume} {59}},\ \bibinfo {pages} {195} (\bibinfo {year} {1985})}\BibitemShut
  {NoStop}%
\bibitem [{\citenamefont {Randeria}(1995)}]{randeria}%
  \BibitemOpen
  \bibfield  {author} {\bibinfo {author} {\bibfnamefont {M.}~\bibnamefont
  {Randeria}},\ }\bibfield  {title} {\bibinfo {title} {\emph {{Crossover from
  BCS Theory to Bose--Einstein Condensation}}},\ }in\ \href {\doibase
  10.1017/CBO9780511524240.017} {\emph {\bibinfo {booktitle} {Bose--Einstein
  Condens.}}},\ \bibinfo {editor} {edited by\ \bibinfo {editor} {\bibfnamefont
  {A.}~\bibnamefont {Griffin}}, \bibinfo {editor} {\bibfnamefont
  {D.}~\bibnamefont {Snoke}}, \ and\ \bibinfo {editor} {\bibfnamefont
  {S.}~\bibnamefont {Stringari}}}\ (\bibinfo  {publisher} {Cambridge University
  Press},\ \bibinfo {address} {Cambridge},\ \bibinfo {year} {1995})\ p.\
  \bibinfo {pages} {355}\BibitemShut {NoStop}%
\bibitem [{\citenamefont {Keeling}\ \emph {et~al.}(2004)\citenamefont
  {Keeling}, \citenamefont {Eastham}, \citenamefont {Szyma\'nska},\ and\
  \citenamefont {Littlewood}}]{Keeling2004}%
  \BibitemOpen
  \bibfield  {author} {\bibinfo {author} {\bibfnamefont {J.}~\bibnamefont
  {Keeling}}, \bibinfo {author} {\bibfnamefont {P.~R.}\ \bibnamefont
  {Eastham}}, \bibinfo {author} {\bibfnamefont {M.~H.}\ \bibnamefont
  {Szyma\'nska}}, \ and\ \bibinfo {author} {\bibfnamefont {P.~B.}\ \bibnamefont
  {Littlewood}},\ }\bibfield  {title} {\bibinfo {title} {\emph {{Polariton
  Condensation with Localized Excitons and Propagating Photons}}},\ }\href
  {\doibase 10.1103/PhysRevLett.93.226403} {\bibfield  {journal} {\bibinfo
  {journal} {Phys. Rev. Lett.}\ }\textbf {\bibinfo {volume} {93}},\ \bibinfo
  {pages} {226403} (\bibinfo {year} {2004})}\BibitemShut {NoStop}%
\bibitem [{\citenamefont {Keeling}\ \emph {et~al.}(2005)\citenamefont
  {Keeling}, \citenamefont {Eastham}, \citenamefont {Szymanska},\ and\
  \citenamefont {Littlewood}}]{Keeling2005}%
  \BibitemOpen
  \bibfield  {author} {\bibinfo {author} {\bibfnamefont {J.}~\bibnamefont
  {Keeling}}, \bibinfo {author} {\bibfnamefont {P.~R.}\ \bibnamefont
  {Eastham}}, \bibinfo {author} {\bibfnamefont {M.~H.}\ \bibnamefont
  {Szymanska}}, \ and\ \bibinfo {author} {\bibfnamefont {P.~B.}\ \bibnamefont
  {Littlewood}},\ }\bibfield  {title} {\bibinfo {title} {\emph {{BCS-BEC
  crossover in a system of microcavity polaritons}}},\ }\href {\doibase
  10.1103/PhysRevB.72.115320} {\bibfield  {journal} {\bibinfo  {journal} {Phys.
  Rev. B}\ }\textbf {\bibinfo {volume} {72}},\ \bibinfo {pages} {115320}
  (\bibinfo {year} {2005})}\BibitemShut {NoStop}%
\bibitem [{Note1()}]{Note1}%
  \BibitemOpen
  \bibinfo {note} {We assume molecules are confined to a two-dimensional plane,
  so there is no dependence on position along the cavity axis.}\BibitemShut
  {Stop}%
\bibitem [{\citenamefont {Eizner}\ \emph {et~al.}(2019)\citenamefont {Eizner},
  \citenamefont {Mart{\'\i}nez-Mart{\'\i}nez}, \citenamefont {Yuen-Zhou},\ and\
  \citenamefont {K{\'e}na-Cohen}}]{eizner2019inverting}%
  \BibitemOpen
  \bibfield  {author} {\bibinfo {author} {\bibfnamefont {E.}~\bibnamefont
  {Eizner}}, \bibinfo {author} {\bibfnamefont {L.~A.}\ \bibnamefont
  {Mart{\'\i}nez-Mart{\'\i}nez}}, \bibinfo {author} {\bibfnamefont
  {J.}~\bibnamefont {Yuen-Zhou}}, \ and\ \bibinfo {author} {\bibfnamefont
  {S.}~\bibnamefont {K{\'e}na-Cohen}},\ }\bibfield  {title} {\bibinfo {title}
  {\emph {Inverting Singlet and Triplet Excited States using Strong
  Light-Matter Coupling}},\ }\href@noop {} {\bibfield  {journal} {\bibinfo
  {journal} {arXiv:1903.09251}\ } (\bibinfo {year} {2019})}\BibitemShut
  {NoStop}%
\bibitem [{\citenamefont {Stone}\ and\ \citenamefont
  {Goldbart}(2009)}]{stone2009mathematics}%
  \BibitemOpen
  \bibfield  {author} {\bibinfo {author} {\bibfnamefont {M.}~\bibnamefont
  {Stone}}\ and\ \bibinfo {author} {\bibfnamefont {P.}~\bibnamefont
  {Goldbart}},\ }\href@noop {} {\emph {\bibinfo {title} {Mathematics for
  physics: a guided tour for graduate students}}}\ (\bibinfo  {publisher}
  {Cambridge University Press},\ \bibinfo {address} {Cambridge},\ \bibinfo
  {year} {2009})\BibitemShut {NoStop}%
\bibitem [{\citenamefont {Arnardottir}\ \emph {et~al.}(2020)\citenamefont
  {Arnardottir}, \citenamefont {Moilanen}, \citenamefont {Strashko},
  \citenamefont {T\"orm\"a},\ and\ \citenamefont {Keeling}}]{SM}%
  \BibitemOpen
  \bibfield  {author} {\bibinfo {author} {\bibfnamefont {K.~B.}\ \bibnamefont
  {Arnardottir}}, \bibinfo {author} {\bibfnamefont {A.~J.}\ \bibnamefont
  {Moilanen}}, \bibinfo {author} {\bibfnamefont {A.}~\bibnamefont {Strashko}},
  \bibinfo {author} {\bibfnamefont {P.}~\bibnamefont {T\"orm\"a}}, \ and\
  \bibinfo {author} {\bibfnamefont {J.}~\bibnamefont {Keeling}},\ }\href@noop
  {} {\bibinfo {title} {\emph {Supplemental Material}}} (\bibinfo {year}
  {2020}),\ \bibinfo {note} {see supplemental material for derivation of the
  cumulant equations, results showing how molecule number affects accuracy of
  mean field theory, details of the high $k$ lasing at large pump power, and
  details of calculation of the photoluminescence spectrum.}\BibitemShut
  {Stop}%
\bibitem [{\citenamefont {Gardiner}(2009)}]{gardiner2009stochastic}%
  \BibitemOpen
  \bibfield  {author} {\bibinfo {author} {\bibfnamefont {C.}~\bibnamefont
  {Gardiner}},\ }\href {https://books.google.co.uk/books?id=otg3PQAACAAJ}
  {\emph {\bibinfo {title} {Stochastic Methods: A Handbook for the Natural and
  Social Sciences}}},\ Springer Series in Synergetics\ (\bibinfo  {publisher}
  {Springer Berlin Heidelberg},\ \bibinfo {year} {2009})\BibitemShut {NoStop}%
\bibitem [{\citenamefont {Kirton}\ and\ \citenamefont
  {Keeling}(2017)}]{Kirton2017}%
  \BibitemOpen
  \bibfield  {author} {\bibinfo {author} {\bibfnamefont {P.}~\bibnamefont
  {Kirton}}\ and\ \bibinfo {author} {\bibfnamefont {J.}~\bibnamefont
  {Keeling}},\ }\bibfield  {title} {\bibinfo {title} {\emph {{Suppressing and
  restoring the Dicke superradiance transition by dephasing and decay}}},\
  }\href {\doibase 10.1103/PhysRevLett.118.123602} {\bibfield  {journal}
  {\bibinfo  {journal} {Phys. Rev. Lett.}\ }\textbf {\bibinfo {volume} {118}},\
  \bibinfo {pages} {123602} (\bibinfo {year} {2017})}\BibitemShut {NoStop}%
\bibitem [{\citenamefont {Kirton}\ and\ \citenamefont
  {Keeling}(2018)}]{Kirton2018}%
  \BibitemOpen
  \bibfield  {author} {\bibinfo {author} {\bibfnamefont {P.}~\bibnamefont
  {Kirton}}\ and\ \bibinfo {author} {\bibfnamefont {J.}~\bibnamefont
  {Keeling}},\ }\bibfield  {title} {\bibinfo {title} {\emph {{Superradiant and
  lasing states in driven-dissipative Dicke models}}},\ }\href {\doibase
  10.1088/1367-2630/aaa11d} {\bibfield  {journal} {\bibinfo  {journal} {New J.
  Phys.}\ }\textbf {\bibinfo {volume} {20}},\ \bibinfo {pages} {015009}
  (\bibinfo {year} {2018})}\BibitemShut {NoStop}%
\bibitem [{\citenamefont {Sánchez-Barquilla}\ \emph
  {et~al.}(2020)\citenamefont {Sánchez-Barquilla}, \citenamefont {Silva},\
  and\ \citenamefont {Feist}}]{SanchezBarquilla2020}%
  \BibitemOpen
  \bibfield  {author} {\bibinfo {author} {\bibfnamefont {M.}~\bibnamefont
  {Sánchez-Barquilla}}, \bibinfo {author} {\bibfnamefont {R.~E.~F.}\
  \bibnamefont {Silva}}, \ and\ \bibinfo {author} {\bibfnamefont
  {J.}~\bibnamefont {Feist}},\ }\bibfield  {title} {\bibinfo {title} {\emph
  {Cumulant expansion for the treatment of light–matter interactions in
  arbitrary material structures}},\ }\href {\doibase 10.1063/1.5138937}
  {\bibfield  {journal} {\bibinfo  {journal} {The Journal of Chemical Physics}\
  }\textbf {\bibinfo {volume} {152}},\ \bibinfo {pages} {034108} (\bibinfo
  {year} {2020})}\BibitemShut {NoStop}%
\bibitem [{Note2()}]{Note2}%
  \BibitemOpen
  \bibinfo {note} {The assumption of single-mode operation used in Eq.~(\ref
  {eq:eomn}--\ref {eq:eomd}) will break down in this coexistence region,
  however this region is very narrow, so we do not expect significant
  deviations to arise.}\BibitemShut {Stop}%
\bibitem [{\citenamefont {{\'{C}}wik}\ \emph {et~al.}(2016)\citenamefont
  {{\'{C}}wik}, \citenamefont {Kirton}, \citenamefont {{De Liberato}},\ and\
  \citenamefont {Keeling}}]{cwik2016excitonic}%
  \BibitemOpen
  \bibfield  {author} {\bibinfo {author} {\bibfnamefont {J.~A.}\ \bibnamefont
  {{\'{C}}wik}}, \bibinfo {author} {\bibfnamefont {P.}~\bibnamefont {Kirton}},
  \bibinfo {author} {\bibfnamefont {S.}~\bibnamefont {{De Liberato}}}, \ and\
  \bibinfo {author} {\bibfnamefont {J.}~\bibnamefont {Keeling}},\ }\bibfield
  {title} {\bibinfo {title} {\emph {{Excitonic spectral features in strongly
  coupled organic polaritons}}},\ }\href {\doibase 10.1103/PhysRevA.93.033840}
  {\bibfield  {journal} {\bibinfo  {journal} {Phys. Rev. A}\ }\textbf {\bibinfo
  {volume} {93}},\ \bibinfo {pages} {033840} (\bibinfo {year}
  {2016})}\BibitemShut {NoStop}%
\bibitem [{\citenamefont {Zeb}\ \emph {et~al.}(2017)\citenamefont {Zeb},
  \citenamefont {Kirton},\ and\ \citenamefont {Keeling}}]{zeb2017exact}%
  \BibitemOpen
  \bibfield  {author} {\bibinfo {author} {\bibfnamefont {M.~A.}\ \bibnamefont
  {Zeb}}, \bibinfo {author} {\bibfnamefont {P.~G.}\ \bibnamefont {Kirton}}, \
  and\ \bibinfo {author} {\bibfnamefont {J.}~\bibnamefont {Keeling}},\
  }\bibfield  {title} {\bibinfo {title} {\emph {{Exact states and spectra of
  vibrationally dressed polaritons}}},\ }\href {\doibase
  10.1021/acsphotonics.7b00916} {\bibfield  {journal} {\bibinfo  {journal} {ACS
  Photonics}\ }\textbf {\bibinfo {volume} {5}},\ \bibinfo {pages} {249}
  (\bibinfo {year} {2017})}\BibitemShut {NoStop}%
\bibitem [{\citenamefont {Breuer}\ and\ \citenamefont
  {Petruccione}(2002)}]{Breuer2002}%
  \BibitemOpen
  \bibfield  {author} {\bibinfo {author} {\bibfnamefont {H.-P.}\ \bibnamefont
  {Breuer}}\ and\ \bibinfo {author} {\bibfnamefont {F.}~\bibnamefont
  {Petruccione}},\ }\href@noop {} {\emph {\bibinfo {title} {{The Theory of Open
  Quantum Systems}}}}\ (\bibinfo  {publisher} {Oxford University Press},\
  \bibinfo {address} {Oxford},\ \bibinfo {year} {2002})\BibitemShut {NoStop}%
\end{thebibliography}
\end{document}